\documentclass[namedreferences]{solarphysics}
\usepackage[hyperref,optionalrh,showbiblabels]{spr-sola-addons} % For Solar Physics 
\usepackage{longtable}
\usepackage{graphicx}   			% For eps figures, newer & more powerfull
\usepackage{array}					% For Rugged Alignment5

  \let\iint\relax

  \let\iiint\relax

\usepackage{amssymb}			% Useful mathematical symbols
\usepackage{amsmath}			% Useful mathematical expressions
\usepackage{color}       			% For color text: \color command
\usepackage{breakurl}        % For breaking URLs easily trough lines
\usepackage[T1]{fontenc}

\usepackage{afterpage}
\newcommand{\etal}{et al.}
\newcommand{\be}{\begin{equation}}
\newcommand{\ee}{\end{equation}}
\newcommand{\beq}{\begin{eqnarray}}
\newcommand{\eeq}{\end{eqnarray}}
\newcommand{\RSUN}{$R_{\odot}$}
\newcommand{\DG}{\ensuremath{^\circ}}
\newcommand{\bl}{} %   
\newcommand{\CDOT}	{\times}
\newcommand{\WIND}		{\emph{Wind}/WAVES}
\newcommand{\IVI}		{{$\text{IV}_\text{IP}$}}
\newcommand{\HL}		{\,-\,halo}
\newcolumntype{L}[1]{>{\raggedright\let\newline\\\arraybackslash\hspace{0pt}}p{#1}}
\newcolumntype{C}[1]{>{\centering\let\newline\\\arraybackslash\hspace{0pt}}p{#1}}  
\newcolumntype{R}[1]{>{\raggedleft\let\newline\\\arraybackslash\hspace{0pt}}p{#1}} 
\newcommand{\apj}  {Astrophys. J.}
\newcommand{\apjl}  {ApJ}
\newcommand{\apjs}  {Astrophys. J. Suppl.}
\newcommand{\aap}  {Astron. Astrophys.}
\newcommand{\aapr}  {Astron. Astrophys. Rev.}

\newcommand{\baas}  {Bull. American Astron. Soc.}

\newcommand{\icarus}  {Icarus}

\newcommand{\solphys}  {Sol. Phys.}

\newcommand{\ssr}  {Space Sci. Rev.}

\newcommand{\grl}  {Geophys. Res. Lett.}

\newcommand{\jgr}  {J. Geophys. Res.}

\newcommand{\TotalEvents}{48}
\newcommand{\Compact}{45}
\newcommand{\CompactX}{19}
\newcommand{\CompactM}{21}
\newcommand{\CompactMX}{40}
\newcommand{\CompactC}{five}
\newcommand{\CompactSlow}{11} % Compact with CME<1000 Km/s (600-900) range
\newcommand{\CompactFast}{32} % Compact with CME<1000 Km/s (600-900) range
\newcommand{\Extended}{three}
\newcommand{\NRHobs}{17}
\newcommand{\IIint}{36}
\newcommand{\PreCondition}{43} % All of them !!!!!!!!!!!!!!! More or Less 
\begin{document}
\begin{article}
\begin{opening}
\title{Interplanetary Type IV Bursts}
%-------------------------------------------------
%% Authors Names
%
\author{A.~\surname{Hillaris}$^{1}$, C.~\surname{Bouratzis}$^{1}$,  A.~\surname{Nindos}$^{2}$}
%-------------------------------------------------
% Runningheads
\runningauthor{Hillaris \etal}
\runningtitle{Interplanetary Type IV Bursts}
%-------------------------------------------------
%% Affilations
\institute{$^{1}$ Section of Astronomy, Astrophysics and Mechanics, Department of Physics,University of Athens, 15783 Athens, Greece{\bl, ahilaris@phys.uoa.g, kbouratz@phys.uoa.gr}\\
           $^{2}$ Section of Astro-geophysics, Department of Physics, University of Ioannina, 45110 Ioannina, Greece, {\bl anindos@cc.uoi.gr}\\}
%-------------------------------------------------
\begin{abstract}
{ }
%CONTEXT{}
{\bl We study} the characteristics of moving  type IV radio bursts which extend to the hectometric wavelengths ({\bl interplanetary type IV or type \IVI~bursts}) and their relationship with energetic phenomena on the Sun. %AIM
Our dataset {\bl comprises \TotalEvents ~interplanetary} type IV bursts observed {\bl with}  the  {\bl Radio and Plasma Wave Investigation    (WAVES) instrument onboard Wind}  in the 13.825 MHz--20 {\bl kHz} frequency range. The dynamic spectra of  the {\bl \emph{Radio Solar Telescope Network} (RSTN), the \emph{Nan\c cay Decametric Array} (DAM), the \emph{Appareil de Routine pour le Traitement et l' Enregistrement Magnetique de l' Information Spectral}  (ARTEMIS-IV), the Culgoora, Hiraiso  and Izmiran \emph{Radio Spectrographs}} were used to track the evolution of the events in the low {\bl corona. These} were supplemented with {\bl soft X-ray (SXR) flux-measurements} from the {\bl \emph{Geostationary Operational Environmental Satellite} (GOES) and coronal mass ejections (CME)}  data from the {\bl \emph{Large Angle and Spectroscopic Coronagraph} (LASCO) onboard the \emph{Solar and Heliospheric Observatory} (SOHO).} Positional information {\bl of}  the coronal bursts {\bl was} obtained by the {\bl \emph{Nan\c cay radioheliograph} (NRH).} 
We examined the relationship of the type IV events with coronal radio bursts, CMEs and SXR flares. %Method
The majority of the events (\Compact) were characterized as {\bl compact}; their duration was on average 106 {\bl minutes}. This type of events {\bl was}, mostly, associated with {\bl M- and X-class flares (\CompactMX~out of  \Compact)} and fast CMEs;  \CompactFast~of  these events had CMEs faster than {\bl 1000 km s$^{-1}$}. Furthermore, in \PreCondition ~{\bl compact}~events the CME was, possibly, {\bl subjected to  reduced aerodynamic drag} as it was propagating in the wake of a previous CME.
 A minority (\Extended) of long lived type \IVI~bursts  was detected, with {\bl duration} from  960 {\bl minutes} to 115 hours. These events are referred to as {\bl extended  or long duration} and   {\bl appear to replenish}  their energetic electron content, possibly from electrons escaping from the corresponding coronal type IV bursts. The latter  were found to persist on the disk, for tens of hours to days. Prominent among them was  the unusual {\bl interplanetary type IV burst of 18-23 May 2002,} which is the longest event in the \WIND~ catalog. {\bl  The \Extended~ extended} events were, usually, accompanied by a number of flares, of  GOES class C in their majority, and  of CMEs, many of which were  slow and narrow.%Results

%-------------------------------------------------
\end{abstract}
%-------------------------------------------------
%% Keywords
\end{opening}
%-------------------------------------------------
\section{Introduction}\label{Intro}

 Solar metric radio bursts provide a unique diagnostic of the development of flare/coronal mass ejection (CME) events in the low corona. Their onset and evolution, is accompanied by energetic-particle acceleration and injection into interplanetary {\bl (IP)} space as well as shocks \citep[see \emph{\bl e.g.} {\bl the} review by~][]{Pick08, Nindos08}. Their signatures at  metric to kilometric wavelengths  trace disturbances propagating from the low corona to {\bl the} interplanetary space.

Three types of nonthermal radio bursts are associated with the {\bl CME/flare} Events \citep{Sakurai74, white2007solar, Gopalswamy2011}: 

\begin{itemize}

\item{Bursts of the {\bl type III family. They}  are produced by energetic electrons accelerated in the Sun and traversing the solar atmosphere,  along coronal magnetic lines rooted in {\bl its surface.} In open field lines, they may escape {\bl into the} interplanetary space (see, for example, {\bl Figure}. 1 of  \citet{Klein08}, also \citet{2015_Alissandrakis}). {\bl In the dynamic spectra, these standard} type III bursts appear as fast drifting bands ($df/fdt\approx1.0s^{-1}$). When trapped in closed magnetic structures, they eventually turn {\bl toward the Sun,} resulting in inverted U- or J-shaped bursts {\bl in} the dynamic spectra (hence  {\bl type-U} or J bursts of the type III family).  Often, in flare/CME events, the transition from the {\bl type-U} and J bursts to the typical type III mark the restructuring, or opening, of the magnetic field lines as originally confined,  energetic electrons gain access to open magnetic lines. An example of dynamic radio spectra showing this  transition from U and J-type bursts to a {\bl standard} type III {\bl burst}  at the beginning of the 17 January 2005 event can be found in \citet{Hillaris2011short}.  In the hectometric and kilometric regime, long duration {\bl storms} of individual type III bursts \citep{Fainberg70,FainbergStone1970SoPh, FainbergStone1971SoPh, 1983_Bougeret} covering several days \citep[5.4 on average after~][]{1987_Kayser} were recorded. These are distinct from the {\bl hectometric\,--\,kilometric extensions} of type III bursts and are known as {\bl IP storms} and, more often than not, may appear as {\bl storm continua} on the dynamic spectra. The individual type III components of the {\bl IP storms},  \citep[{\bl micro-type III} bursts, after][]{2007_Morioka} are significantly weaker than the typical type III bursts in the same frequency range. The {\bl IP storms} are well associated with active regions \citep{1987_Kayser} yet the {\bl micro-type III} bursts are not accompanied by significant SXR flare activity. This implies the need of a persistent coronal store of suprathermal electrons \citep{1984a_Bougeret, 1984b_Bougeret} supplying this type of activity.}

\item{Type  II {\bl bursts. They} are the radio signatures of  the passage of {\bl a magneto-hydrodynamic (MHD)} shock wave through the tenuous plasma of the solar corona; their radio emission is due to energetic electrons accelerated at the shock front.  It is, in general, accepted that {\bl type} II bursts at decametric and longer wavelengths are driven by CMEs, bow or flank \citep{Vrsnak2008}. At the metric range, on the other hand, they might be also due, apart from CMEs,  to flare blasts \citep{1988_Cane&Reames_a, 2011_Nindos&al, Magdalenic2010, Magdalenic2012} or reconnection outflow {\bl termination-shocks} \citep{Aurass02}.}

\item{Type IV {\bl bursts. They}  are  radio continua  due to the radiation of energetic electrons trapped within magnetic structures and plasmoids. They have been recorded in almost all frequency ranges starting from the microwaves as  type \mbox{IV${\mu}$}  bursts and the  decimetric range as \mbox{\bl type IVdm} \citep{Benz1980}.  In  the metric wavelengths the type IVm bursts are divided into  moving (IVmA or IVM) and  stationary (IVmB). The type IVmB bursts emanate from stationary magnetic structures usually located above active regions or post-eruption arcades behind CMEs \citep{Robinson85, Gopalswamy2011}. The type IVm burst are, sometimes,  referenced as flare continua \citep[FCM when preceding a type IVmA or FCII {\bl when following} a type II burst, see~][]{Robinson78B}. They are also identified as continuum noise storms  \citep[\bl type IVsA and IVsB, corresponding to type IVmA and IVmB, see discussion in~][\bl ~and their Figure 14]{Sakurai74}. The type IVmA  bursts \citep{Boischot1957} appear moving outwards at velocities of the order of 100--1000 km $s^{-1}$ comparable to CME speeds \citep{white2007solar}; they sometimes last more than {\bl ten minutes}.  A number of these  are believed to originate within the densest substructures of CMEs \citep{KleinMouradian02, Bastian01, Aurass99, Bain2014}.  These substructures are, possibly, erupting prominences within the CMEs. The type IVmA burst---CME association was found to increase with the speed of the CME \citep[][\bl ~and references therein]{Gergely1986}. A subset of the moving type IV radio bursts extend, {\bl in} dynamic spectra, to the hectometric wavelengths (frequencies {\bl lower} than $\approx 20$ MHz) and are recorded {\bl with} the {\bl \emph{Radio and Plasma Wave Investigation} (WAVES) instrument onboard \emph{Wind}}; these are {\bl interplanetary type IV} or \IVI~radio bursts.}

\end{itemize}

In this {\bl article} we examine the characteristics and the evolution of  {\bl interplanetary type IV} bursts and their relationship with energetic phenomena on the Sun such as flares and {\bl CMEs. The}  data used are from the \WIND~ receivers and a number of ground based instruments. From the combined {\bl datasets} an extensive table of type \IVI~and associated activity was compiled and is presented  {\bl as supplementary online material (file IPtypeIV.pdf). A detailed description of the table is included in the Appendix.} The question addressed in our {\bl study is twofold. Firstly,}  we examine the association of these bursts with intense flares and fast CMEs, examining whether they may be considered as another aspect of the  {\bl big flare syndrome} first introduced by \citet{1982_Kahler_b}. {\bl Secondly,  we} search for other processes affecting, {\bl totally or partially}, the appearance of this type of radio bursts.
 
This report on  {\bl interplanetary type IV (or Type--\IVI)} bursts  is structured as {\bl follows.} In Section \ref{Obs} we describe the instrumentation and {\bl dataset} used in our study. The data analysis  {\bf is} presented in {\bl Section} \ref{Analysis}~including an overview of selected events  in {\bl Sections} \ref{Over02}, \ref{Over01}, and \ref{Over01B}.  In {\bl Section} \ref{res} we present the characteristics and the evolution of different types of  {\bl type \IVI}~events which, then, are discussed in {\bl Section \ref{discussion}. The} conclusions are presented in the same section.  

\section{Observations and Data Selection} \label{Obs}

The basic data used in this study are dynamic spectra recorded by the R1 and R2 receivers of the \WIND~\citep{Bougeret95short}  in the 20kHz--13.825 MHz frequency range from 1998 to 2012. The  interplanetary type IV bursts selected {\bl were already identified} in the\WIND~{\bl online} catalog\footnote{www.\-lep.gsfc.nasa.gov/waves/data\_products.html}. The observations were complemented by data {\bl in the metric and decametric} wavelengths from the  following ground-based radio observatories: 

\begin{itemize}

\item  {\bl The \emph{Artemis-IV radio-spectrograph\footnote{Appareil de Routine pour le Traitement et l' Enregistrement Magnetique de l' Information Spectral, http://artemis-iv.phys.uoa.gr/}} \citep[][\bl observes in the frequency range 20--650 MHz.]{Caroubalos01short, KontogeorgosShort, Kontogeorgos06short, Kontogeorgos08short}.}  
\item The {\bl \emph{Culgoora  radio-spectrograph}} \citep[][\bl observes in the frequency range 18-1800 MHz.]{Prestage94}.
\item The {\bl \emph{Nan\c cay Decametric Array or DAM\footnote{bass2000.obspm.fr/home.php}}} \citep[][\bl observes in the range 20--75 MHz.]{Boischot1980short,Lecacheux2000}. 
\item {\bl \emph{The Nan\c cay Radioheliograph}} \citep[\bl NRH:][]{Kerdraon97} provides daily, 09:00-15:30 UT, two-dimensional images of the Sun at {\bl ten} frequencies from 450 to 150 MHz  with sub-second time resolution. It was used for positional information of the metric--decametric radio {\bl emission. In this article the  quick-look-style} NRH data from the radio-monitoring site\footnote{http://radio-monitoring.obspm.fr/nrh\_data.php} were used.
\item The {\bl \emph {Hiraiso Radio Spectrograph\footnote{sunbase.nict.go.jp/solar/denpa/index.html}}} \citep[\bl HiRAS:][]{Kondo95} {\bl observes in the frequency range 25--2500 MHz}.
 \item {\bl \emph{The Institute of Terrestrial Magnetism, Ionosphere and Radio Wave Propagation (IZMIRAN) Radio Spectrograph}\footnote{www.izmiran.ru/stp/lars/}} \citep[][]{Gorgutsa2001} {\bl observes in the range 25\,--\,270 MHz.}
\item {\bl \emph{The Radio Solar Telescope Network}\footnote{ftp://ftp.ngdc.noaa.gov/STP/space-weather/solar-data/solar-features/solar-radio/rstn-spectral}} \citep[\bl RSTN:][]{Guidice81} with a number of solar radio observatories at various locations around the world {\bl guarantees full, 24 hours}, coverage.: 
	\begin{itemize}
	\item{Sagamore Hill   at Hamilton, Massachusetts, USA (42\DG 33$^\prime$N 70\DG 49$^\prime$W)}
	\item{Palehua   at Kaena Point, Hawaii (21\DG 24$^\prime$N 158\DG 06$^\prime$W)}
	\item{Holloman   at New Mexico, USA (32\DG 51$^\prime$N 106\DG 06$^\prime$W)}
	\item{Learmonth   {\bl at Western Australia, Australia} (22\DG 13$^\prime$S--114\DG 06$^\prime$E)}
	\item{San Vito dei Normanni, Italy (40\DG 39$^\prime$N 17\DG 42$^\prime$E)}
	\end{itemize}
\noindent These observatories provide dynamic spectra in the {\bl 25\,--\,180} MHz range.
\end{itemize}

Additional {\bl datasets} were used in order to examine the association of the type \IVI~bursts with the evolution of flares and CMEs:
\begin{itemize}
\item{CME data from the {\bl \emph{Large Angle and Spectroscopic Coronagraph} (LASCO) Catalogue } on line\footnote{http://cdaw.gsfc.nasa.gov/CME{$\_$}list}~\citep{Yashiro04,Gopalswamy2009}}
\item{SXR ({\bl \emph{Geostationary Operational Environmental Satellite}, GOES})  characteristics from on line reports\footnote{\mbox{ftp://ftp.ngdc.noaa.gov/STP/space-weather/solar-data/solar-features/solar-flares/x-rays/} and \mbox{https://solarmonitor.org/data}} and light curves\footnote{http://satdat.ngdc.noaa.gov/sem/goes/data/}.}
\item {Images from the \emph{Extreme Ultraviolet Imaging Telescope} \citep[\bl EIT:][]{Delaboudiniere95short} {\bl  onboard the \emph{Solar and Heliospheric Observatory} SOHO. They}  were used in order to provide information on flare positions.}
\end{itemize}

From the \WIND~catalog  all events indicated as bursts of type \IVI~(\TotalEvents~ in total) were selected. Many (\IIint) were accompanied by interplanetary type II shocks. 

A comprehensive list of the interplanetary type IV bursts and the associated activity including, but not restricted to, coronal burst, flare and CME data {\bl is attached as supplementary online material (file IPtypeIV.pdf). A detailed description of the table is included in the Appendix}.   In compiling this catalog, we included information of all the\WIND~type \IVI~bursts,  their associated CMEs and SXR flares in the 1998\,--\,2012 period and the accompanying interplanetary type II and coronal {\bl type} II, IV and III bursts.  The above mentioned CMEs which are thought to drive the type \IVI~bursts are refered to as {\bl main CMEs} in order to distinguish from preceding CMEs along the same path; the latter were included in the table as they may affect the appearace of type \IVI~bursts as discussed in {\bl Sections} \ref{CharCompact} and \ref{discussion}. The selection of the CMEs preceding {\bl the main ejection} along the same path is based on a time interval of about 48 hours before the main CME and whether the sectors (or cones in 3D) defined by position angle {\bl and width} for the main and the preceeding {\bl CME overlap. Occasionally} there are more than one preceding CMEs included in the catalog {\bl because they are all within the 48 hours window and they appear to overlap, in part, with the main CME}.  The overlap criterion is relaxed when one or both the main and the preceding CMEs are {\bl halo CMEs}; in this case  we assume that an overlap is always possible, at least in part.

The catalog format is described, in detail, in  {\bl the appendix.}

\section{Data Analysis}\label{Analysis}

\subsection{Data Processing}\label{DataAnalysis}

The combination of hectometric dynamic spectra by\WIND~with metric and decametric spectra from the {\bl ground-based radio spectrographs} (see {\bl Section} \ref{Obs}) was, firstly, plotted {\bl as a composite dynamic spectrum.} These were found to include an amount of features, mostly groups of type III and II bursts, embedded in a slowly varying background. {\bl Often the continuum background was}  removed by the use  of high--pass filtering on the dynamic spectra (usually differentiation).  This filtering, however, amplifies the high-frequency noises therefore the smoothed differentiation filter of \citet{usui1982} was used. This performs simultaneously a smoothing and differentiation {\bl so it can be} regarded as a low-pass differentiation filter (digital differentiator) appropriate for experimental (noisy) data processing.  

The composite dynamic spectra provide an overview of the evolution of the type \IVI~bursts under study, and of the accompanying radio activity, from the corona to the interplanetary space. On each dynamic spectrum {\bl several time--histories} were superposed:
\begin{itemize}
\item The approximate frequency--time trajectories of the CME {\bl fronts. These} were plotted on the dynamic spectra, using the coronal density model of  \citet{Vrsnak04}, as dashed {\bl lines. The selection details for the model are} presented in {\bl Section} \ref{Density}.  The linear fits to the height--time trajectories of the CME fronts, from the LASCO images, were converted to the frequency--time traces of the fundamental and harmonic plasma emission; the squares mark the measured positions of the CME front.
\item The GOES SXR time--profiles: The  solid black (1.0\,--\,8.0 \AA) and the dotted purple (0.5\,--\,4.0 \AA) curves display the SXR time history describing thermal emission from the {\bl hot flare} plasma.
\end{itemize}
\noindent Of the  \TotalEvents ~type \IVI~bursts of this report \NRHobs ~overlapped, at least partly, with the NRH window of observation. For these the position of {\bl the} coronal  extension (metric type IV burst) of the interplanetary burst was compared to the SXR flare position and the solar sources of the CME {\bl in the EIT images. Their} spatial relationship was, thus, established.  This was obtained by a combination  of NRH radio contours overlayed on the EIT 195 \AA ~difference image and combined with the LASCO difference image. {\bl Two categories of type \IVI~bursts were found:}

\begin{figure}
\centering
\includegraphics[trim=16cm 0cm  14cm 0cm,clip=true,width=\textwidth]{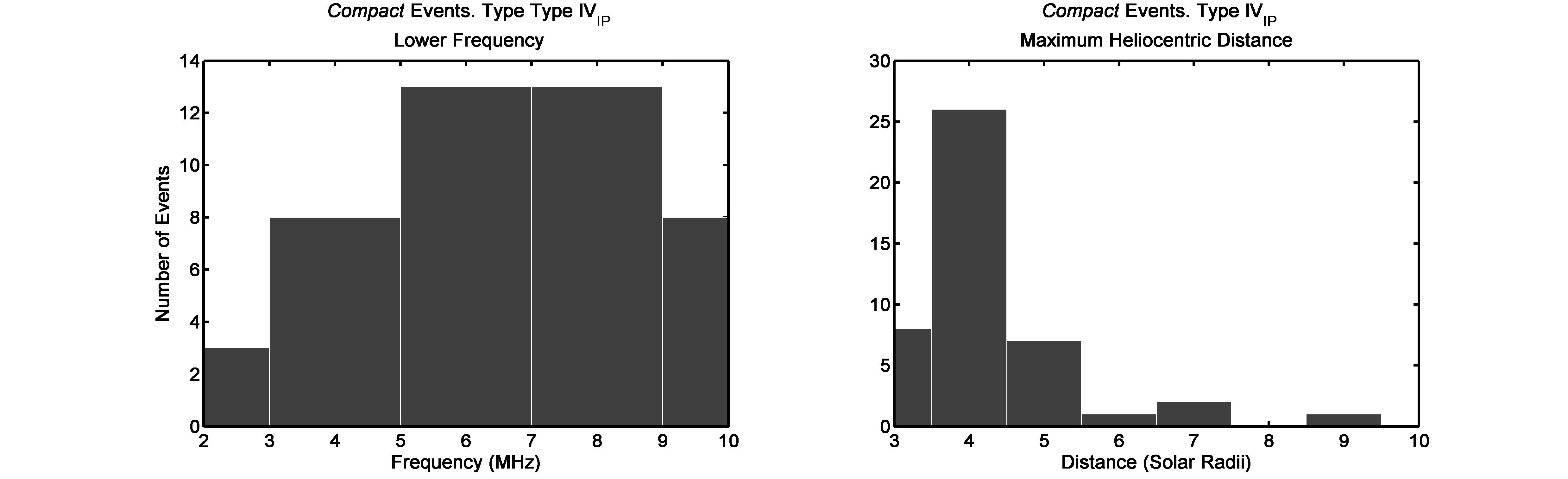} 
  \caption{{\bl Left panel: Distribution} of the low frequency limit of the {\bl compact} interplanetary type IV bursts. {\bl Rigth panel: The} corresponding heliocentric distance of these type IV bursts based on the low frequency limit and the model depedent calculations described in {\bl Section} \ref{Density}.}
 \label{LowFreq}
 \end{figure}

\begin{figure}
\centering
\includegraphics[trim=16cm 0cm  09cm 0cm,clip=true,width=\textwidth]{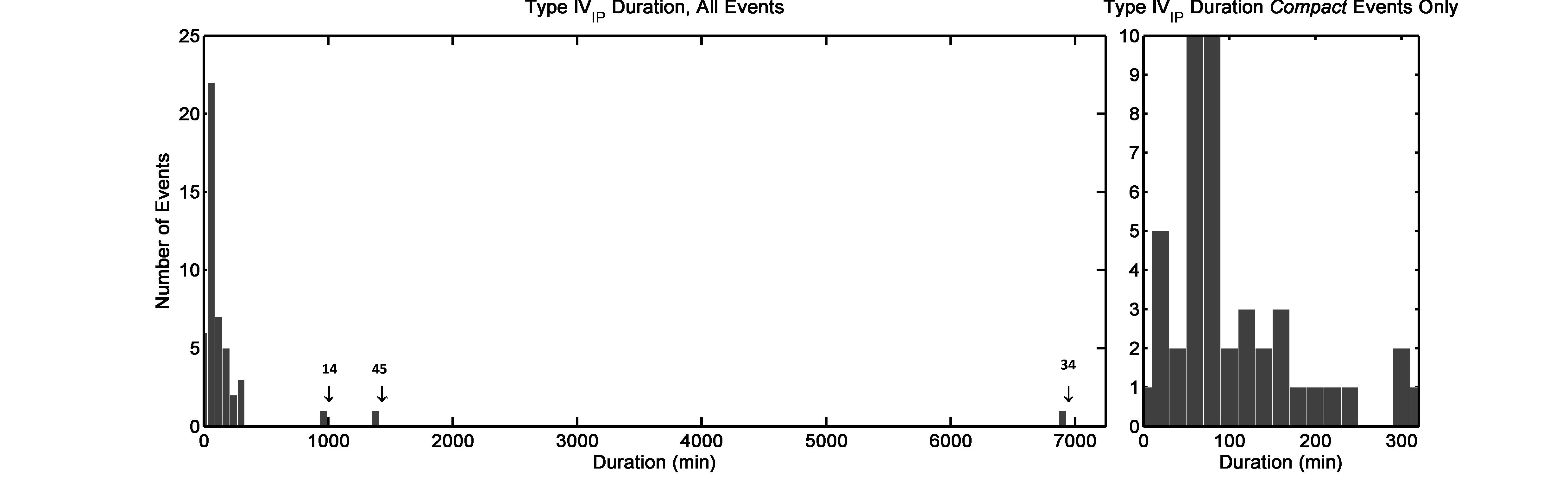} 
  \caption{ Distribution of the {\bl duration} of the interplanetary type IV bursts. {\bl Left panel: All (\TotalEvents) events; the \Extended~extended events are pointed} with {\bl arrows. Their catalog numbers are shown above them. Right panel:} Histogram {\bl including only the \Compact~compact events}}
 \label{Duration}
 \end{figure}

\begin{itemize}
\item [\bl Compact type \IVI~bursts.]{These were, mostly, associated with {\bl M- and X-class} flares and fast CMEs; their duration was on average 100 (\mbox{$\pm$11}) {\bl minutes}. Their minimum frequency was in the 10\,--\,2 MHz range which corresponds to 3\,--\,10 \RSUN~heliocentric {\bl distances. The} distribution of the low frequency limits and the corresponding distaces, derived from the calculations in {\bl Section} \ref{Density}, are exhibited {\bl in Figure} \ref{LowFreq}. In total, \Compact ~of these events were found in the\WIND~lists. An example is presented in  {\bl Section}  \ref{Over02}.}
\item [\bl Long Duration or \emph{Extended} type \IVI~bursts.]{They represent a small minority of \Extended ~events with durations from  960 {\bl minutes} to 115 hours. Their morphology was found to be less uniform than that of their majority counterparts. Two of them (catalog numbers 34, {\bl 18\,--\,23 May 2002}, and 45, {\bl 27 May 1999,} described in {\bl Sections} \ref{Over01}~and~ \ref{Over01B}) were accompanied by a sequence of small flares and {\bl slow and narrow} CMEs with an occasional medium or large flare within the sequence.  {\bl Another event (number 14, 17 January 17 2005, presented in Section}~\ref{Over01C}) originated from {\bl a fast CME and large flare, but extended far beyond the duration of a compact event.}}
\end{itemize}
\noindent  The distribution of the interplanetary {\bl type IV bursts duration is presented in Figure} \ref{Duration} for all the events of the study. {\bl The \Extended ~extended} events are the outliers of  the histogram on the left panel. The distribution of the {\bl duration of the compact events} is also presented, separately, on the right panel of {\bl Figure} \ref{Duration}.  In {\bl Section} \ref{res} the differences between the characteristics of the {\bl extended and the compactt} events, apart from {\bl their} duration, are examined and discussed.
%-------------------------------------------------
%-------------------------------------------------
%-------------------------------------------------
%-------------------------------------------------
\subsection{Coronal Density--Height Model Selection}\label{Density}
As plasma emission depends on electron density, which in turn may be converted to coronal height using density models, we may calculate the radio source heights and speeds from {\bl the} dynamic spectra. The establishment of a correspondence between frequency of observation--coronal height and frequency drift rate--radial speed is affected by ambiguities introduced by the variation of the ambient medium properties. These may be the result of {\bl the} burst exciter propagation within undisturbed plasma, {\bl over-dense or under-dense} structure or CME after-flows \citep[\bl see~][for a detailed discussion on model selection]{Pohjolainen07,Pohjolainen08}. 

The density model of  \citet{Vrsnak04}.
%---------------------------------------------------------------------------------------------------
 \begin{displaymath}
 \frac{n}{10^8 \; \rm cm^{-3}}=15.45 \left( \frac{R_\odot}{R} \right)^{16}  
                                                    + 3.165 \left( \frac{R_\odot}{R} \right)^6  
                                                    + 1.0 \left( \frac{R_\odot}{R} \right)^4  
                                                    + 0.0033 \left( \frac{R_\odot}{R} \right) ^2,  
 \end{displaymath}

\noindent which describes well the coronal density behavior in the large range of distances from low corona to interplanetary space was used for the conversion of  the linear fits to the height-time trajectories of the LASCO CME fronts  to frequency-time tracks on the composite dynamic spectra.

%-------------------------------------------------
 \begin{figure}
\centering
\includegraphics[trim=5cm 4cm  3cm 2cm,clip=true,width=0.9\textwidth]{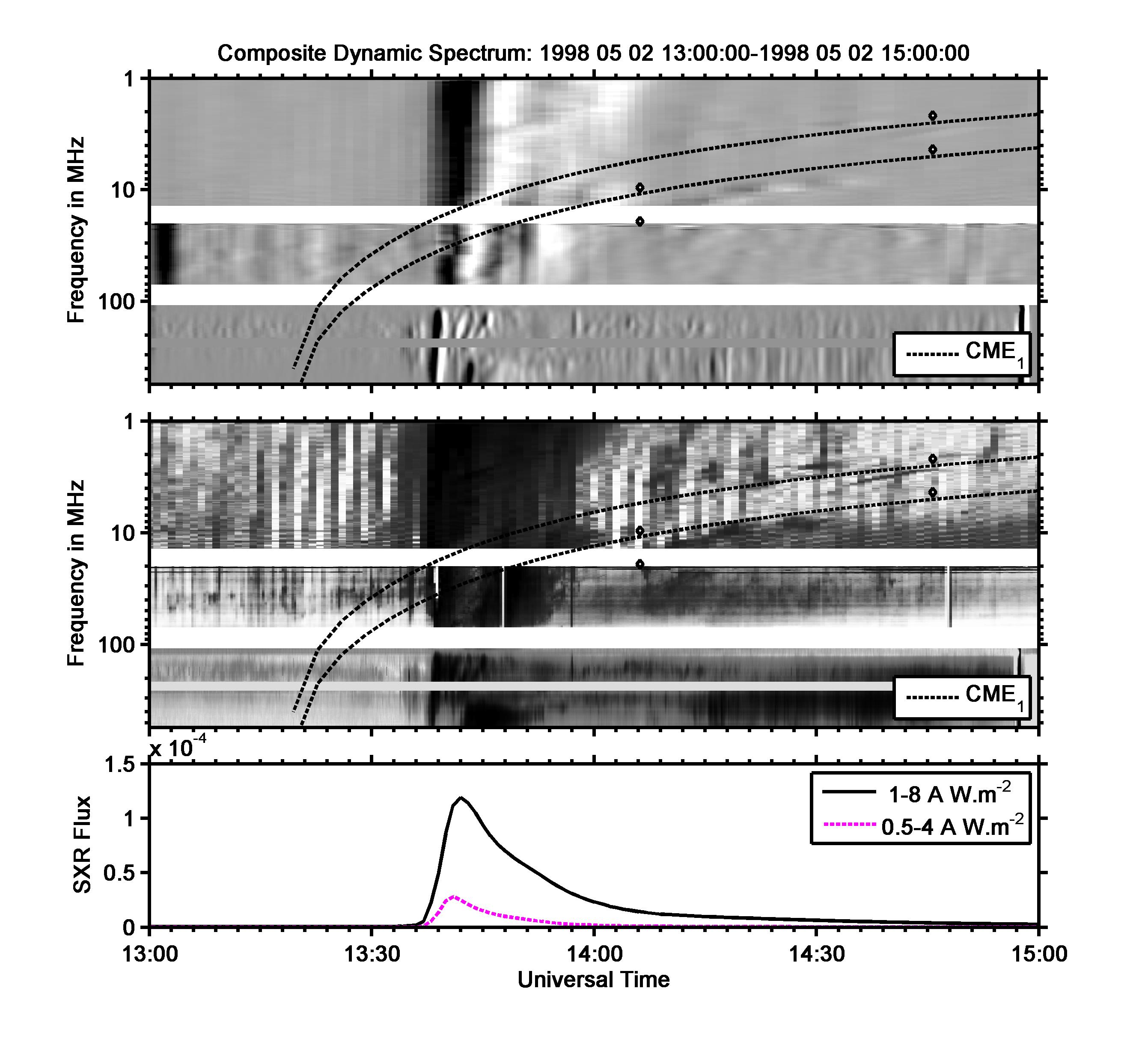} 
  \caption{{\bl 02 May 1998 event.}. Top pannel: {\bl \WIND~\emph{ARTEMIS-IV}} differential  spectrum (inverse grey scale). Middle Panel: Dynamic ({\bl intensity,} inverse grey scale) spectrum. The frequency-time plots derived from the linear fits to the front trajectory of the associated CME and an empirical density model (see {\bl Section} \ref{Density}) for {\bl the} fundamental  and harmonic (dashed curves) plasma emission are overlayed on the spectra. Bottom Panel: The profiles of GOES SXR 1-8 \AA~({\bl solid-black line}) and 0.4-4 \AA~({\bl dashed-red line}) flux.}
 \label{DynSpec19980502}
 \end{figure}
%-------------------------------------------------
 \begin{figure}
\centering
\begin{tabular}{lr}
\includegraphics[width=0.4\textwidth]{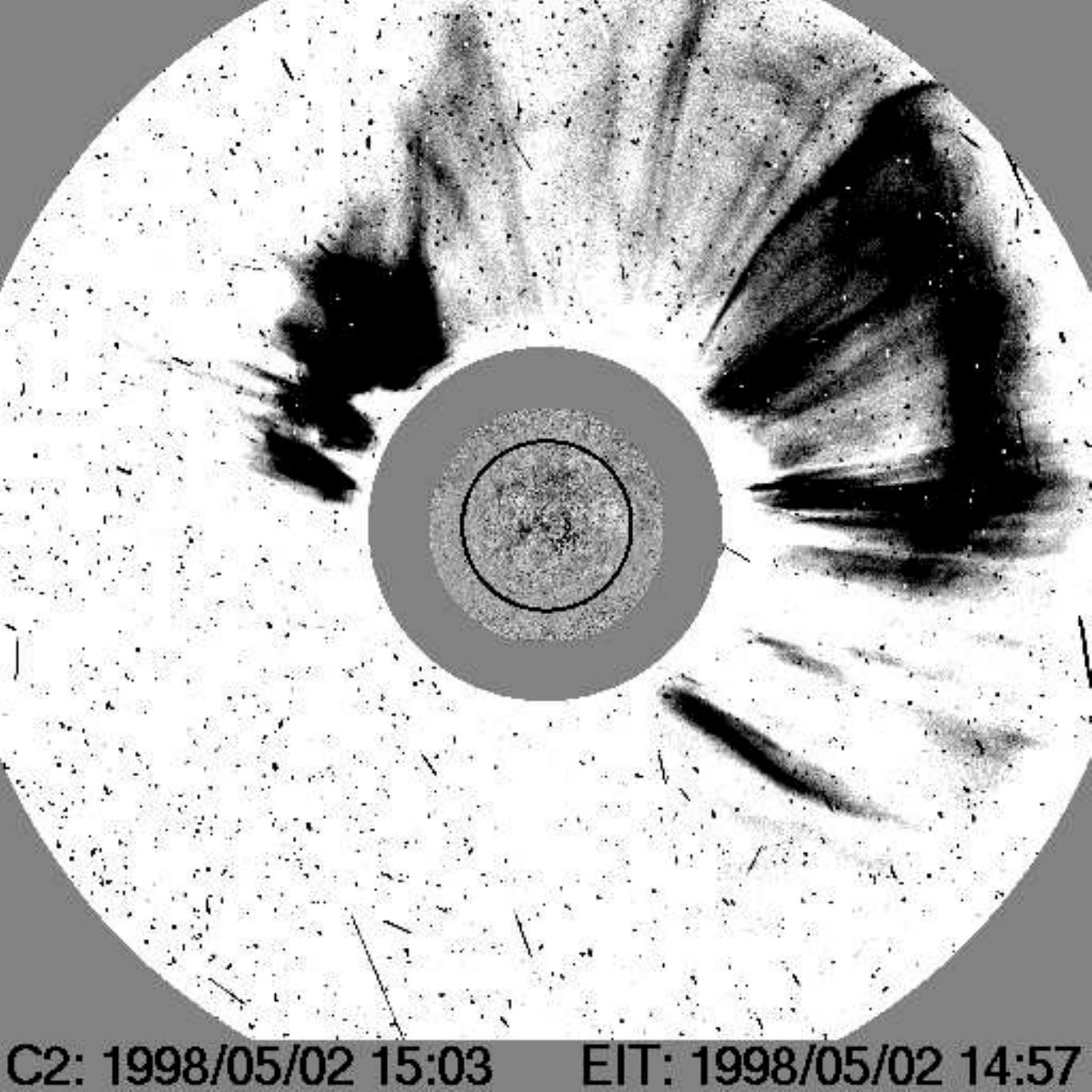} &
\includegraphics[width=0.45\textwidth]{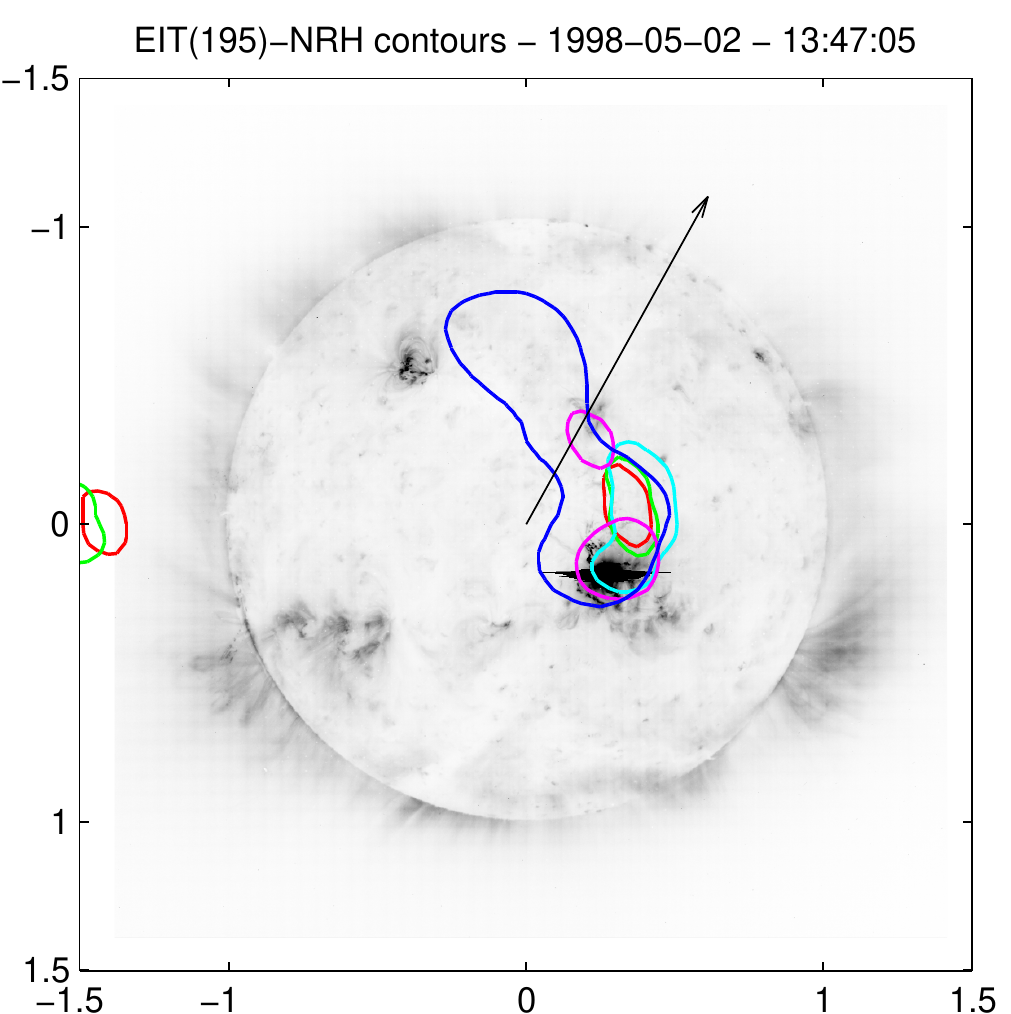} \\
 \end{tabular}
  \caption{Left: LASCO {\bl and}  \mbox{EIT 195 \AA}~ running difference frames of the {\bl 02 May 1998} 14:06 UT CME (inverse gray scale). Right: NRH half power contours (at 432 (red), 410 (green), 327 (cyan), 236 (magenta), and 164 (blue) MHz). The contours were recorded {\bl at} successive times {\bl starting with 432 MHz at 13:47:05 UT to 164 MHz at 13:51:53 UT, thus tracing the outward motion of the type IV burst. The arrow indicates the halo CME measured position angle from the LASCO catalog.}}
 \label{NRH19980502}
 \end{figure}
%-------------------------------------------------
\subsection{Overview  of the 2 May 1998  {\bl Compact} Event Evolution}\label{Over02}
{The  {\bl 2 May 1998 compact event} (catalog number 47) is typical of its class.  It has drawn considerable attention due to the large number of instruments that have observed it, including \WIND, SOHO/LASCO, EIT, NRH, and several radio spectrographs.}  It is reported in  a number of articles which focus, mainly, on the solar surface {\bl magnetic waves} \citep[][]{1999_Zharkova&Kosovichev}, the pre--CME launch activity \citep{Pohjolainen1999a} and the on disk development of the CME \citep{Pohjolainen2001short}. 

The interplanetary type IV event (see composite spectrum {\bl in} Figure \ref{DynSpec19980502}) starts at  14:10 UT on {\bl 02 May and lasts until} 15:40 UT of the same day in the frequency range {\bl 8-14 MHz. An} interplanetary type II burst was recorded from  14:25--14:50  UT in the 5--3 MHz range the \WIND. In the catalog it is described as {\bl a narrowband wisp}  yet it is well associated with the front of the CME. Another type II, without apparent association to the CME front,  appears in the 400--6 MHz range, recorded by  the{\bl  \emph{ARTEMIS--IV}, the  \emph{Nan\c cay Decametric Array}}  (DAM) and the  \WIND~from {\bl 13:30\,--\,13:46 UT. It} exhibits {\bl a multiple} band structure and was first reported by \citet{Pohjolainen2001short}. The high frequency extension of the type \IVI~was recorded by the DAM and the {\bl \emph{ARTEMIS--IV} radio spectrographs} and extends above the 500 MHz \citep[see also~][~their Figure 8]{Pohjolainen2001short}.  This activity is accompanied by an \mbox{\bl X1.1/3B-class}  flare from {\bl active region} AR 8210 at heliographic coordinates {\bl S15W15;} the flare started at 13:31 and ended at 13:51 UT peaking at 13:42 UT. 

The NRH {\bl records} at  432 and 164 MHz indicate that the type IV continuum appears over AR 8210 and, on the 164 MHz images, starts moving northwards at 13:34 {\bl UT. This} is consistent with the movement of a rather fast, 938 km $s^{-1}$ halo CME (first view at 14:06 UT, back extrapolated lift--off at 13:07 UT) with measured position angle 331\DG (see Figure \ref{NRH19980502}). This CME appears driving the fundamental {\bl and} harmonic pair, mentioned in the previous paragraph. As regards the CME path, there are two preceding {\bl halo} CMEs {\bl on 02 May 1998} at 05:32 UT and on {\bl 01 May 1998} at 23:40 UT. 

The broadband dynamic spectra and the NRH images indicate that the {\bl compact} interplanetary  type IV burst  is associated with an {\bl X-class} flare and a fast halo CME. The latter propagates in the wake of a previous halo CME which was launched approximately eight and a half hours before. This example represents the combined effects of an intense flare {\bl and} fast CME event with the CME propagating within a {\bl low-drag} region due to the passage of a previous CME.

\subsection{\bl Overview of the 18--22 May 2002 Extended Event Evolution}\label{Over01}
\begin{figure}[]
\centering
\includegraphics[trim=13.5cm 3.5cm  11.5cm 3.5cm,clip=true,width=\textwidth]{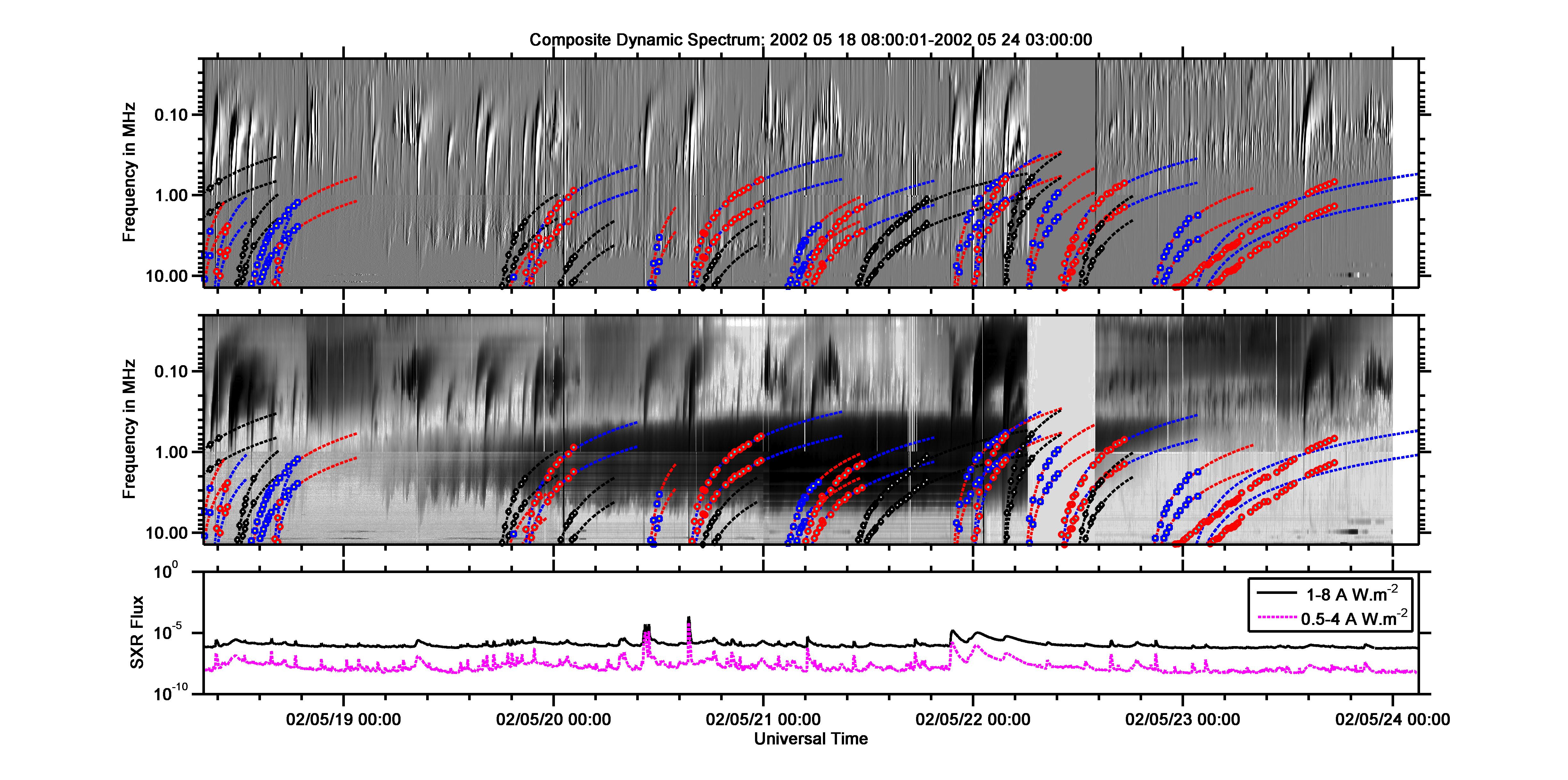} 
  \caption{\bl 18-22 May 2002 event from 18:08:00 UT to 24:03:00 UT. Top pannel:   \WIND~differential  spectrum (inverse grey scale). Middle Panel: Dynamic (intensity, inverse grey scale) spectrum. The frequency-time plots derived from the linear fits to the front trajectory of the associated CMEs and the density model presented in  Section \ref{Density}  for  the fundamental  and harmonic (dashed curves) plasma emission are overlayed on the spectra. In this case  the CME trajectories are of different colors. Bottom Panel: The profiles of GOES SXR 1-8 \AA~(solid-black line) and 0.4-4 \AA~(dashed-red line) flux The time is in written in the format date, hours, and minutes.}
\label{DynSpec20020518}
 \end{figure}

\begin{figure}
\centering
\includegraphics[trim=4.0cm 4cm 9cm 3.5cm,clip=true,width=0.9\textwidth]{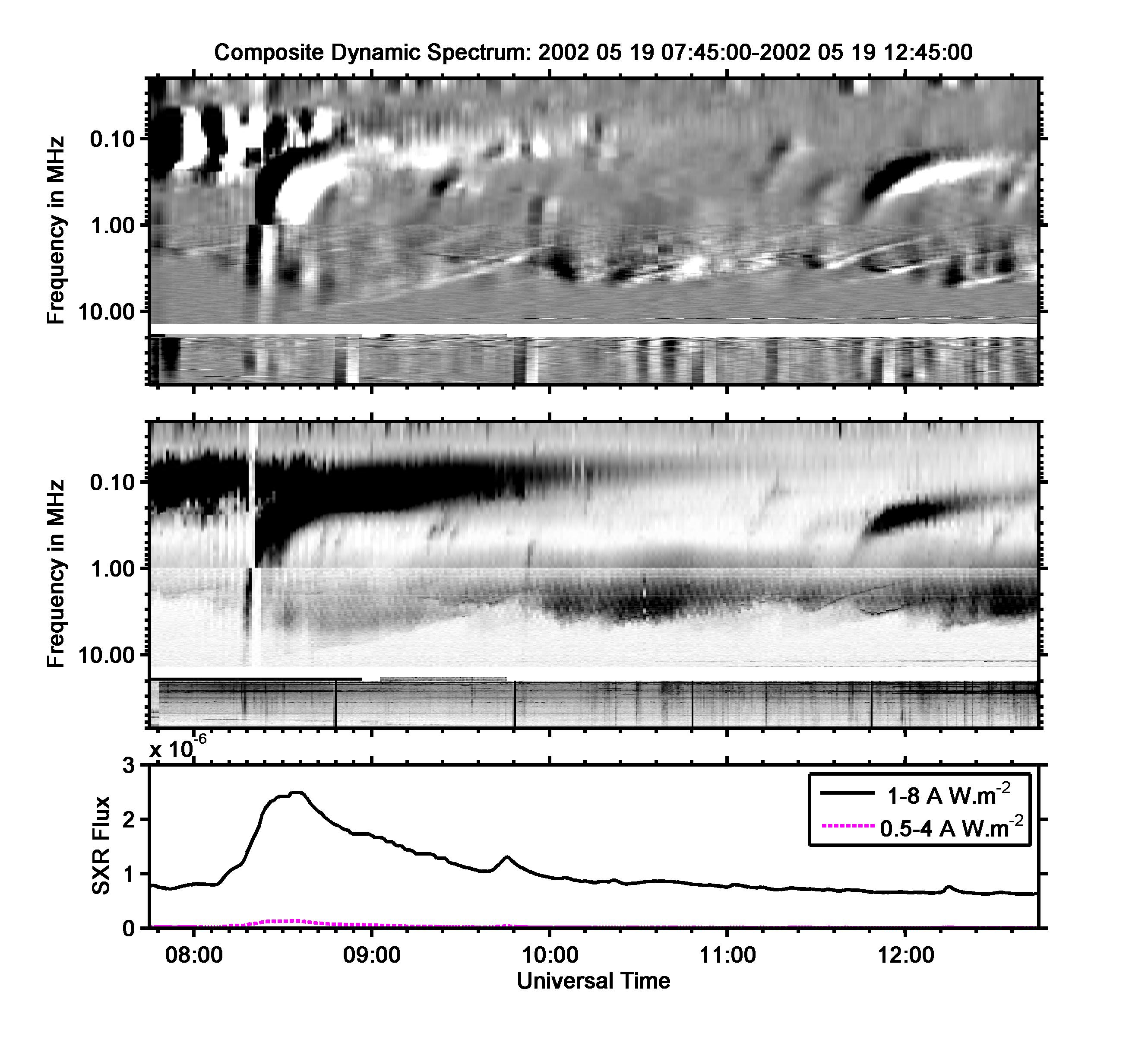} 
    \caption{\bl 18--22 May 2002 event with details on 19 May from 07:45 UT to 12:45 UT. The dynamic spectra are combined records from  \WIND,  DAM and extended in the 0.02--70 MHz range. Top pannel:  Differential  spectrum (inverse grey scale). Middle Panel: Dynamic (intensity, inverse grey scale) spectrum.  Bottom Panel: The profiles of GOES SXR 1-8 \AA~(solid-black line) and 0.4-4 \AA~(dashed-red line) flux }
\label{DynSpec20020518A}
 \end{figure}

{The type \IVI ~event starts at 09:00 UT on {\bl 18 May} and lasts until 04:00 UT {\bl on 23 May} in the frequency range {\bl 0.3\,--\,9MHz} (catalog number {\bl 34). It} was the only such event observed by\WIND ~in its years of operation \citep{Reiner2006}. An interplanetary type II burst was recorded from the 22 May at 04:10 UT until the 23 May 10:10 {\bl UT} in the 0.5--0.03 MHz range.}  The\WIND~dynamic and differential spectra, with the CME front trajectories overlayed, and the SXR flux are exhibited in {\bl Figure} \ref{DynSpec20020518}; details in the period {\bl 07:45\,--\,12:45 UT on 19 May} are presented in {\bl Figure} \ref{DynSpec20020518A}. This event was briefly reported by \citet{Gopalswamy2005} who, based on polarization measurements, considers a hectometric storm continuum and not a type \IVI ~burst as reported in the\WIND~catalog. 
%-------------------------------------------------
%-------------------------------------------------

\begin{figure}
\centering
\includegraphics[trim=19.5cm 18.0cm  12.0cm 8.0cm,clip=true,width=\textwidth]{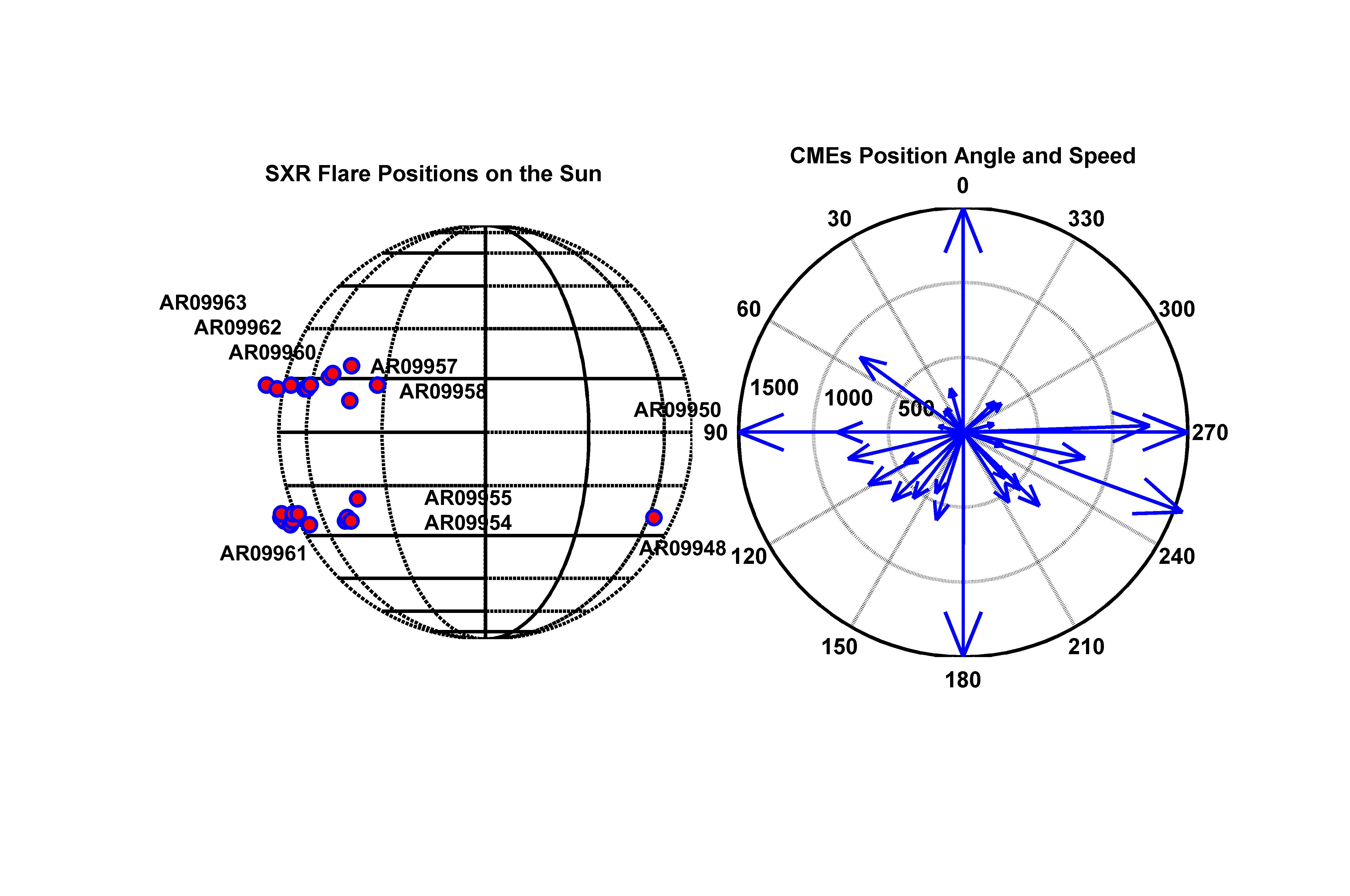} 
  \caption{Left: Positions of the SXR flares (purple dots with blue border) and active-regions.  Right: CME position angles {\bl and CME speeds indicated with segments ending in an arrow-head,  The segment length is proportional to the CME speed. The active} region positions {\bl are taken from} the Solar Monitor {\bl (www.solarmonitor.org)}.  The flare and CME positions  extend in the whole {\bl 18\,--\,23 May 2002} period; the AR positions are from the middle of this period on {\bl 20 May}.}
\label{CMEpos20020518}
 \end{figure}

On the solar disk, we see a number of {\bl active-region} complexes in the {\bl 18\,--\,23} May period. When the continuum started, on {\bl 18 May} {\bl AR 9957 (N08E47) was the largest and most complex region on the disk. This region was accompanied by AR 9958 (N04E45) and followed by AR 9960 at N05E74 and AR 9962, AR 9963 which appeared on the 21 May at N15E47 and N17E63. South of this group were AR  9954 (S22E35), AR 9955 (S14E37) and in the western hemisphere AR 9945 (S02W73), AR 9948(S21W20) and AR 9950 (S05W42). On the left panel of  Figure \ref{CMEpos20020518} we present the positions of all SXR flares within the 18--23  period and of the corresponding active regions. }

\begin{figure}
\centering
\includegraphics[trim=12.0cm 6.0cm  10.0cm 2.0cm,clip=true,width=\textwidth]{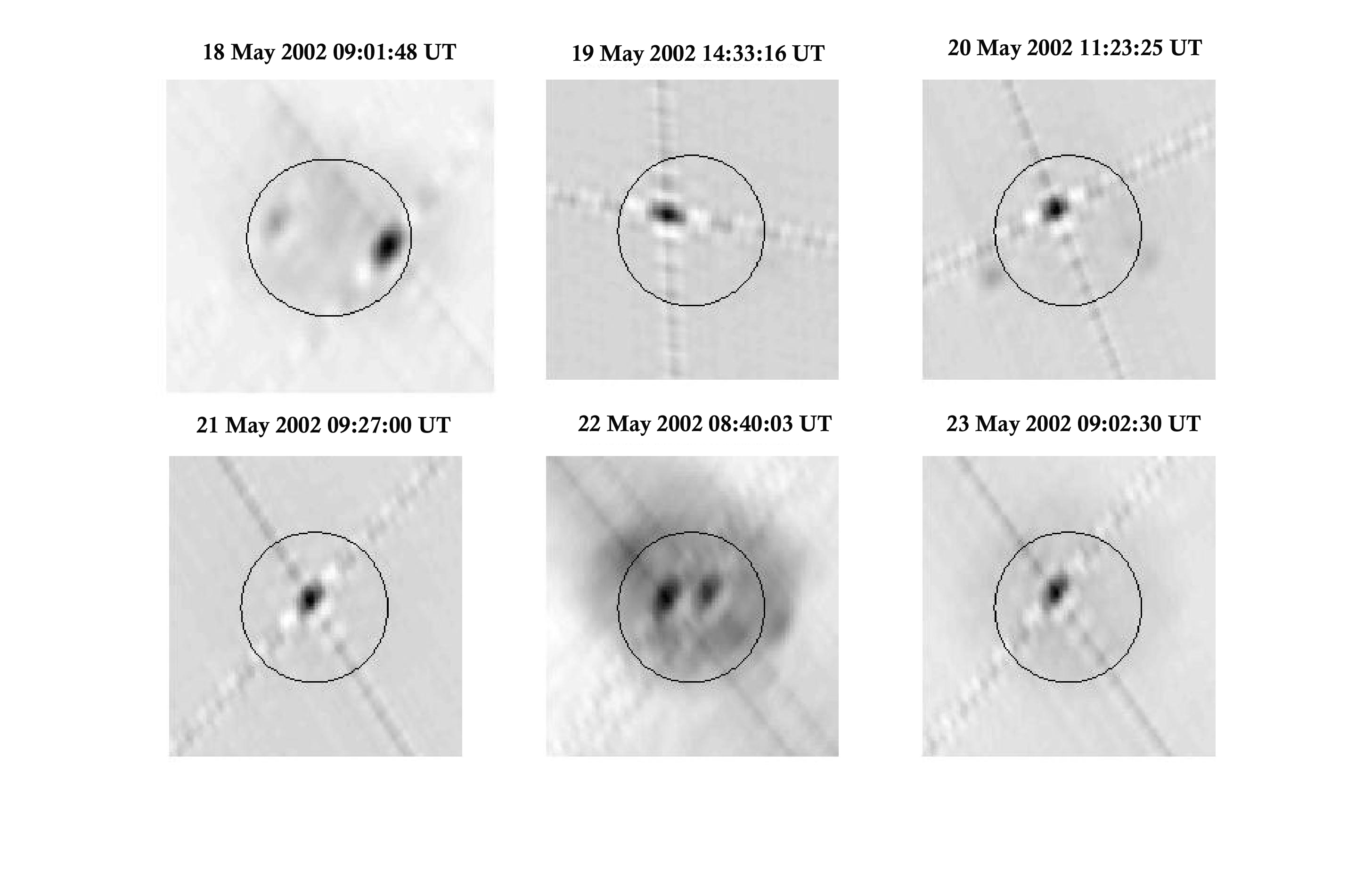} 
  \caption{Positions of the {\bl coronal type} IV bursts in the  {\bl 18--22 May 2002 period}, obtained by the \emph{Nan\c cay Radioheliograph} at 164 MHz.}
\label{NRH20020518}
 \end{figure}

{\bl Throughout} the period of interest {\bl from 18 May 2002 at 02:44 UT to 23 May 2002 at 04:00 UT,} 38 SXR flares were recorded by GOES, positional data were obtained for 33 of them. {\bl The AR 9957, AR 9958, AR 9960, AR 9962, AR 9963 complex, located in the NE quadrant of the disk, gave 10 C-class and one M1.5 SXR flares. The single AR 9961 in the SE quadrant gave 16 flares (including an X2.1 and an M5.0) on 19 May 2002 at 15:54 UT. }Within the same period 24 CMEs were recorded in the LASCO {\bl catalog. The } positional angles indicate that they emerged from all quadrants of the solar disk. This activity was associated with many type III {\bl burst-groups, 2\,--\,3} type II shocks in the metric range {\bl (see the table included as supplementary online material)} and a persistent continuum appearing on the NE quadrant over the AR 9957, AR 9958, AR 9960, AR 9962, AR 9963 group {\bl during the 19\,--\,23}  May {\bl period. The} metric and decametric continuum appears on the SW quadrant, over AR 9948, only on {\bl 18 May (see Figure} \ref{NRH20020518}); on this day it coexists with the persistent continuum (over the AR 9957) mentioned above. 
%-------------------------------------------------
%-------------------------------------------------
%-------------------------------------------------
%-------------------------------------------------

The SXR activity originates, mostly, in the NE (AR 9957, AR 9958, AR 9960, AR 9962, AR 9963 group) and the SE (AR 9961) quadrants; most of the CME position angles indicate {\bl ejections} from the same two quadrants (see {\bl Figure} \ref{CMEpos20020518}, right panel). The position of the type III {\bl bursts, which could be localized, were almost} equally divided between the SE and the NE quadrants but the vast majority appears clearly on the differential spectra to continue into the\WIND~hectometric range at frequencies lower {\bl than}  the  type \IVI. Furthermore the type III and CME activity continues past the end of the interplanetary type IV burst. The metric continuum, on the other hand, appears persistently, in the NRH images, on top of the AR 9957, AR 9958, AR9960, AR 9962, AR 9963 group from the 19 to the 23 May. This implies a {{\bl steady coronal reservoir}} of energetic electrons, which may follow the magnetic lines trailing CMEs originating at the NE quadrant and replenish the electrons of the interplanetary type \IVI ~burst.

%-------------------------------------------------
 \begin{figure}
\centering
\includegraphics[trim=12cm 3.5cm  12cm 2cm,clip=true,width=\textwidth]{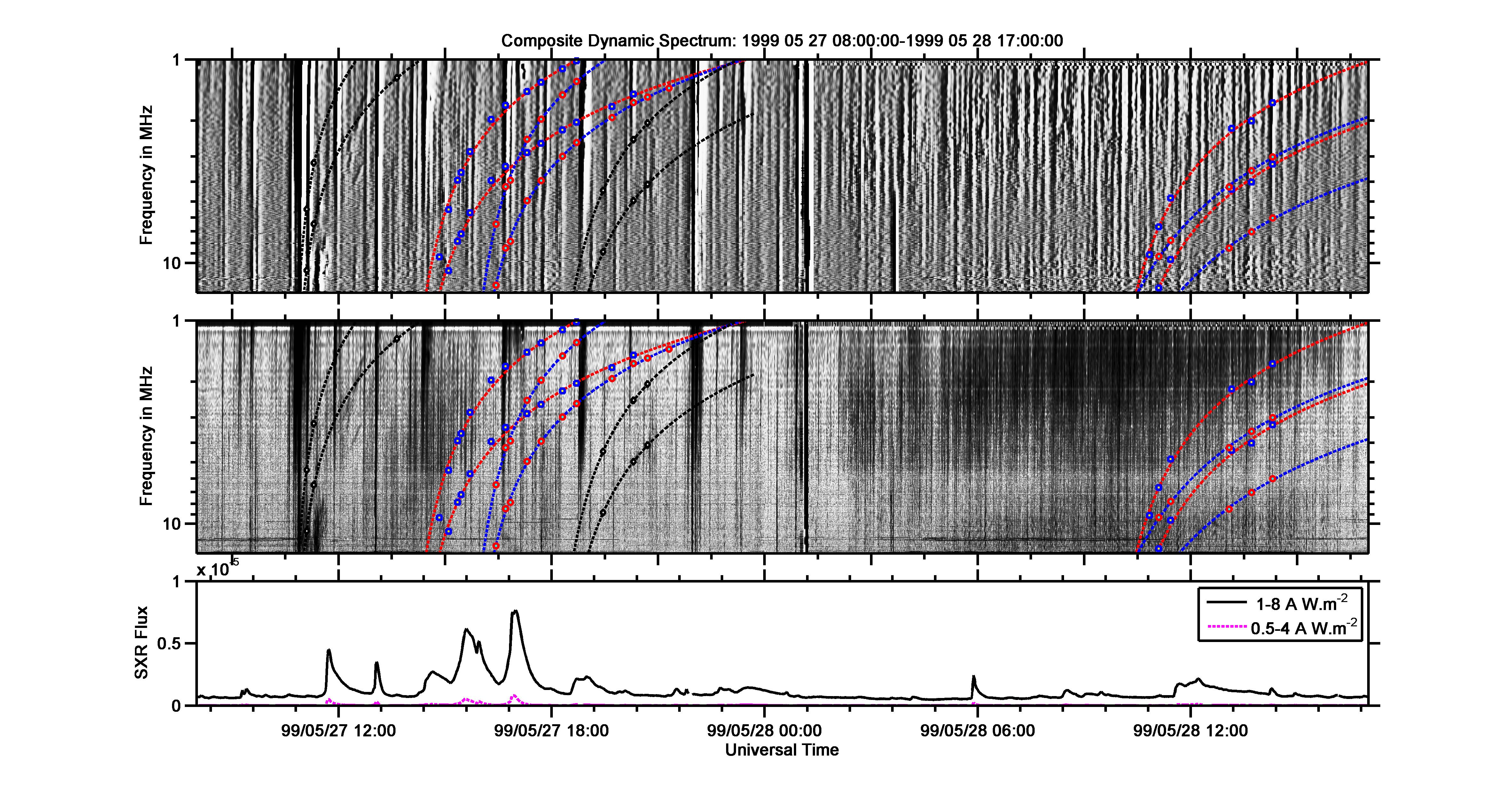} 
  \caption{\bl 27--28 May 1999 event in the period form 27 May 08:00 UT to 28 May 17:00 UT. Top pannel: \WIND~differential spectrum (inverse gray scale). Middle Panel: Dynamic (intensity)  spectrum. The frequency-time plots  derived from the linear fits to the front trajectory of the associated CME and an empirical density model (see  Section \ref{Density}) for the fundamental  and harmonic (dashed curves) plasma emission are overlayed on the spectra. Bottom Panel: The profiles of GOES SXR 1\,--\,8 \AA~(solid-black line) and 0.4-4 \AA~(dashed-red line) flux. The time is in the format date hour and minutes.}
  \label{DynSpec19990527}
 \end{figure}
%-------------------------------------------------
 \begin{figure}
\centering
\includegraphics[trim=6.5cm 4cm  5.5cm 3.5cm,clip=true,width=\textwidth]{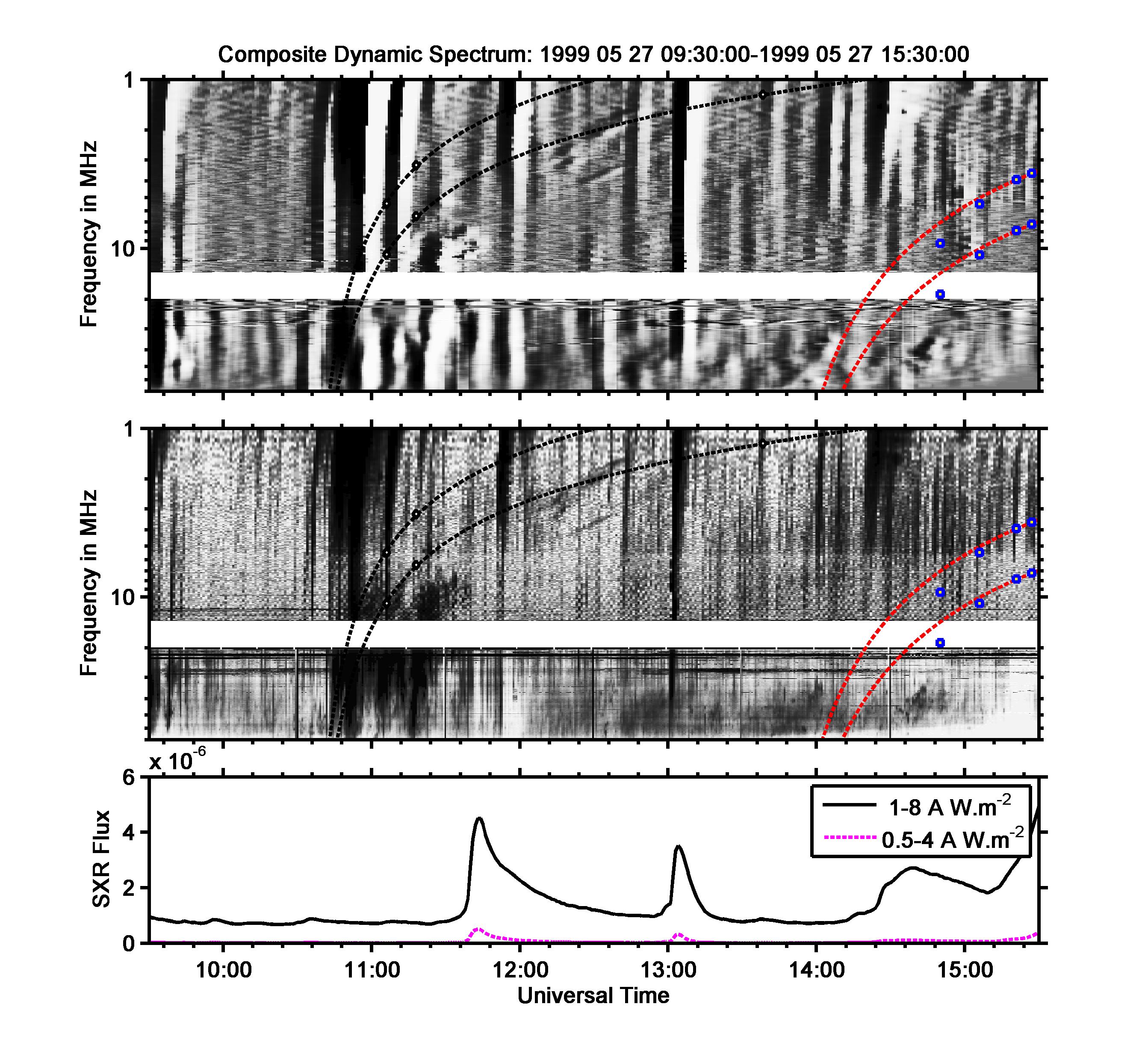} 
  \caption{\bl 27--28  1999 event. The panels show the event in  the 27 May 09:30--15:30 UT period. The dynamic spectra are combined records from  the\WIND~and the DAM  in the 1--70 MHz range. This section of the type \IVI ~starts above 70 MHz and extends to 1 MHz.Top pannel:differential spectrum (inverse gray scale). Middle Panel: Dynamic (intensity)  spectrum. The frequency-time plots  from the linear fits to the front trajectory of the associated CME and the density model presented in Section \ref{Density}) for the fundamental  and harmonic (dashed curves) plasma emission are overlayed on the spectra.Bottom Panel: The profiles of GOES SXR 1\,--\,8 \AA~(solid-black line) and 0.4-4 \AA~(dashed-red line) flux.  }
\label{DynSpec19990527A}
 \end{figure}
%-------------------------------------------------
 \begin{figure}
\centering
\includegraphics[trim=6.0cm 4cm  6cm 3.5cm,clip=true,width=\textwidth]{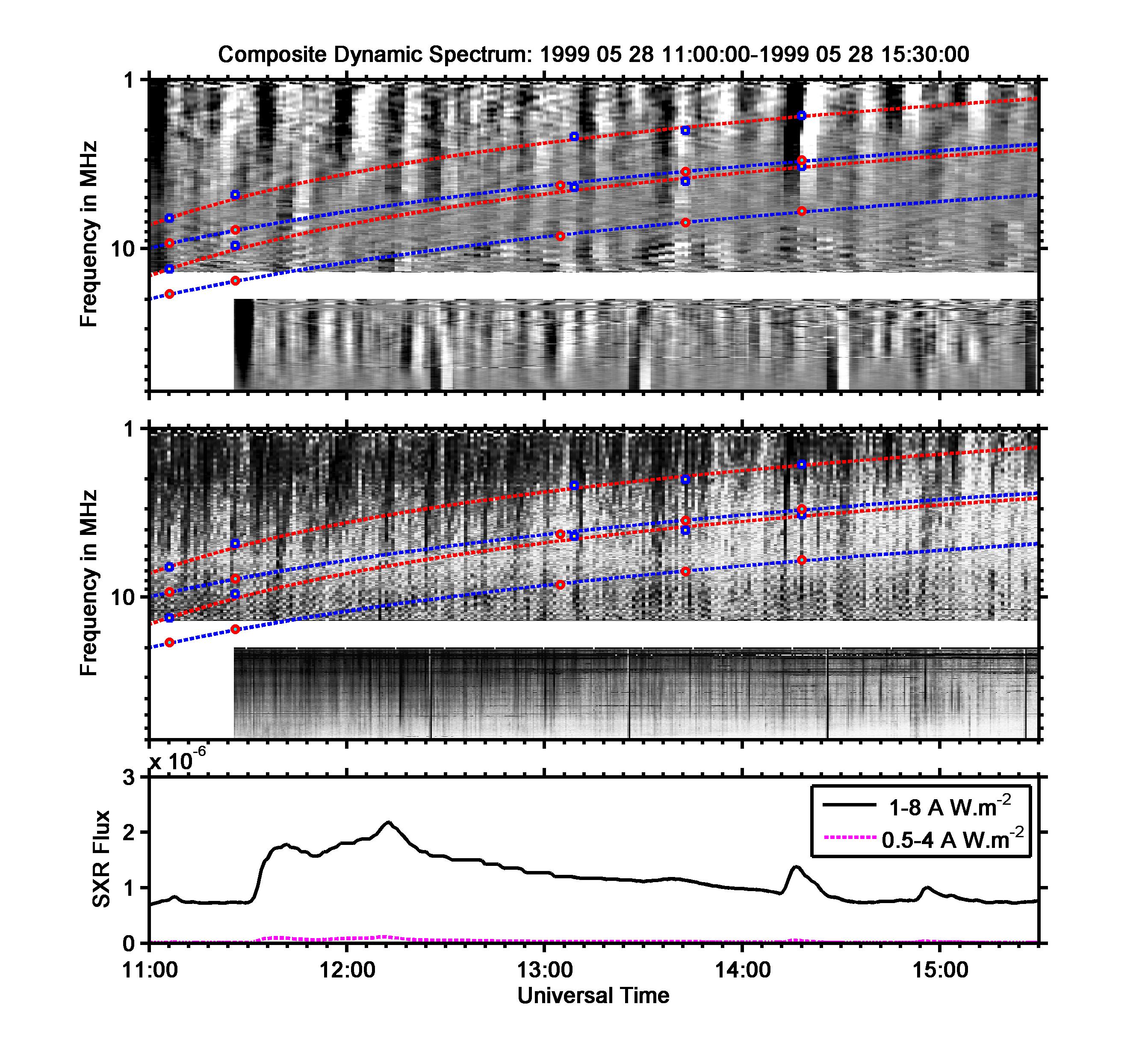} 
  \caption{\bl 27--28  1999 event. The panels show the event in  the 27 May 11:00--15:30 period. The dynamic spectra are combined records from  the\WIND~and the DAM  in the 1--70 MHz range. The type \IVI ~has a coronal extension above 70 MHz; the most prominent part reaches 1 MHz with some parts extending below this frequency.Top pannel:differential spectrum (inverse gray scale). Middle Panel: Dynamic (intensity)  spectrum. The frequency-time plots  from the linear fits to the front trajectory of the associated CME and the density model presented in Section \ref{Density} for the fundamental  and harmonic (dashed curves) plasma emission are overlayed on the spectra.Bottom Panel: The profiles of GOES SXR 1\,--\,8 \AA~(solid-black line) and 0.4-4 \AA~(dashed-red line) flux.  }  
\label{DynSpec19990527B}
 \end{figure}
%-------------------------------------------------
%-------------------------------------------------
 \begin{figure}
\centering
\includegraphics[trim=12.0cm 6.0cm  10.0cm 2.0cm,clip=true,width=0.9\textwidth]{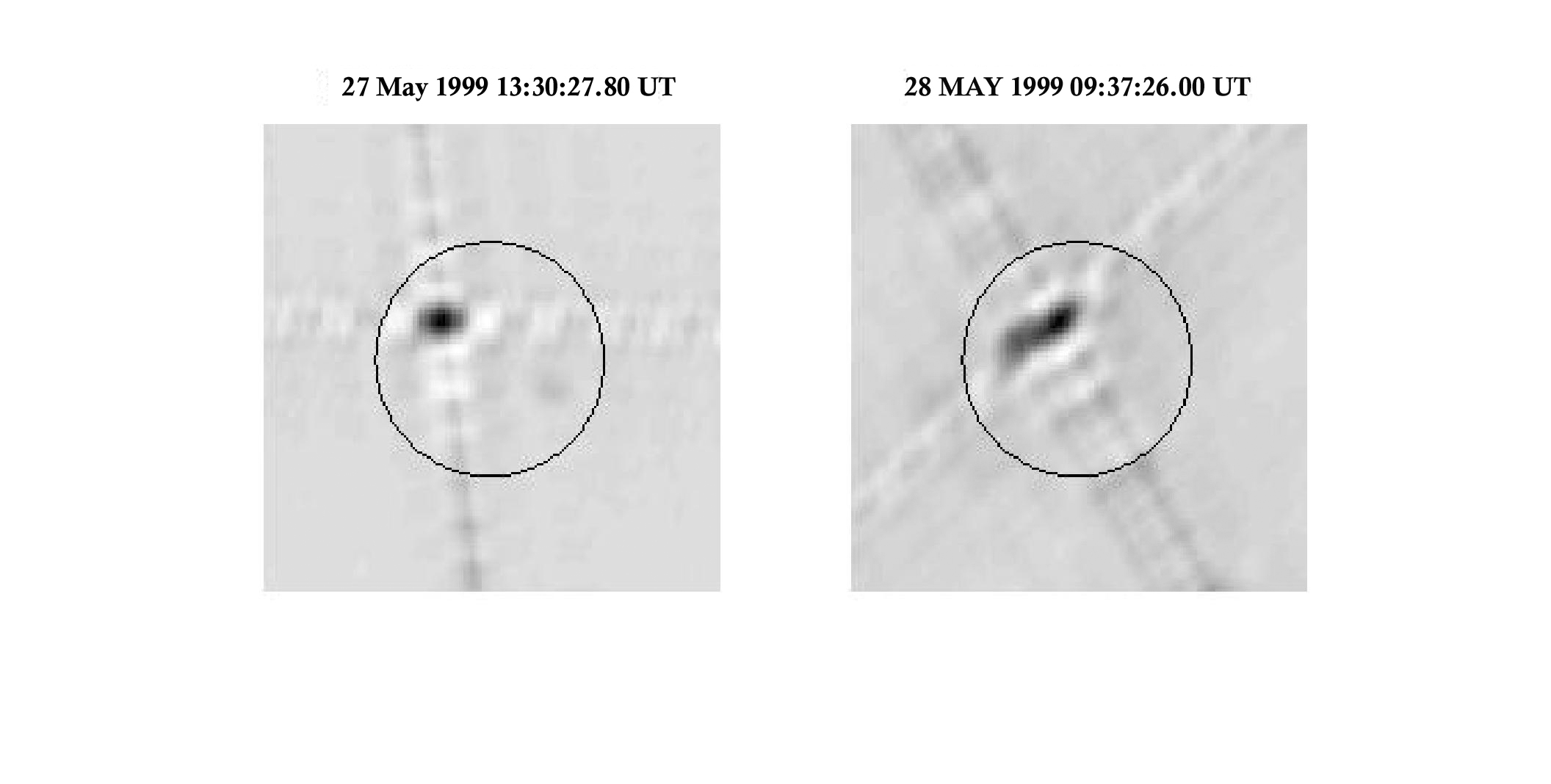} 
  \caption{Positions of the coronal type IV bursts in the 27--28 May 1999 period, obtained by the \emph{Nan\c cay Radioheliograph} at 164 MHz.}
\label{NRH19990527}
 \end{figure}
%-------------------------------------------------

\subsection{\bl Overview of the 27--28 May 1999 Extended Event Evolution}\label{Over01B}
The event starts {\bl on 27 May 1999 at} 10:55 UT and ends {\bl on} 28 May at 15:00 UT (catalogue number 45).{\bl The event is  composed by interplanetary type II/IV bursts}  where both  have coronal extensions.  This event is accompanied by a number of C-class SXR flares and narrow CMEs. An overview including \WIND~ dynamic spectrum, the CME front {\bl trajectories, and the SXR} flux profiles is presented in Figure \ref{DynSpec19990527}. There are only two {\bl wide CMEs, a halo} on {\bl 27 May} at 11:06 UT and a rather wide CME on {\bl 28 May} at 10:26 UT,  which almost mark the start and the end of the event. 

Similar to the {\bl  18--22 May 2002 } event, in {\bl Section} (\ref{Over01}), there is also a persistent coronal type IV, {\bl which appears over AR 8552 (N18E31) on the NRH records.} This region remains active throughout the duration of the interplanetary type IV burst and most of the small SXR {\bl flares  and a} number of type III bursts originate {\bl from it as well.} In Figure \ref{NRH19990527} we present the position of the coronal type IV, on {\bl  27 and 28 May},  in the form of NRH images. There is also a  long  series of type III bursts and groups, which covers the type \IVI~interval. Most of these type III bursts, however, overshoot the type \IVI~continuum so we expect that the main source of energetic electrons is its coronal counterpart  persisting over {\bl AR8552}.

The end of the  {\bl 27--28  May 1999 event} coincides with the wide CME mentioned at  the beginning of the paragraph.

\subsection{\bl Overview of the 17 January 2005 Extended Event Evolution}\label{Over01C}
On 17 January 2005 two fast {\bl CMEs} were recorded in close succession during two distinct episodes of a 3B/X3.8 flare from the AR 10720. The type \IVI ~ burst  started at 10:55 UT on 17 January and ended by {\bl 18 January}  at 02:00 UT {\bl (event catalog number 14), for} an overview of the dynamic spectrum see Figure 3 of \citet{Hillaris2011short}.  The coronal extension of the type \IVI~burst was found to originate from AR10720 and persisted throughout the  duration of its interplanetary counterpart. The type II activity, on the other hand, was restricted to the frequency range below 14 MHz. The  type III groups accompanying the event were found to overshoot the low frequency limit of the type \IVI~burst at least after 08:18 UT. Earlier,  groups of type U bursts and type IV fine structures indicated acceleration and partial trapping of electrons behind the CME front.  

A detailed study of the radio signatures of this event \citep[Table 1 in~][ {\bl that } presents a comprehensive outline of its time evolution]{Hillaris2011short} points towards {\bl possible multiple}  acceleration mechanisms. These include CME associated shocks  in the high corona and the interplanetary space, and also, shock--independent accelerators at low altitudes associated with the type IV continuum behind the CME.  

This event had distinct features of the {\bl compact} class, associated with an intense flare and two fast CMEs, yet its long duration characterizes it as {\bl extended}. Similar to the previous two long duration events, the energetic electrons provided by low corona sources are well  associated with the coronal type IV burst.

\section{Characteristics of All Events}\label{res}
\subsection{Characteristics  of the  \Compact~{\bl Compact}  Events}\label{CharCompact}
Of the \TotalEvents ~type \IVI~bursts of our sample \Compact, classified as {\bl compact}, were found to conform to the {\bl big flare syndrome}  which suggests that, statistically, energetic phenomena are more intense in larger flares, regardless of the detailed physics. In fact \CompactX~were associated {\bl with X-class} flares and \CompactM~{\bl with M-class} with only \CompactC ~events {\bl related to  C-class} flares.This represents a significant deviation with respect to the general GOES SXR flare distribution studied by \citet{2002D_Veronig&al}. In the general case it is expected that $\sim$66\% of the SXR flares {\bl are} of class C with $\sim$9.5\% {\bl of class M} and just $\sim$0.7\%  {\bl of class X.}.  The general case is, also, consistent with the power-law distribution of the peak SXR flux [I], where the probability density, $p(I)\sim I^{-b}$ with $b\approx 2$  \citep[see][]{2012_Aschwanden&Freeland}. 
%-------------------------------------------------
 \begin{figure}
\centering
\includegraphics[trim=8.0cm 2.0cm  6.0cm 4.0cm,clip=true,width=\textwidth]{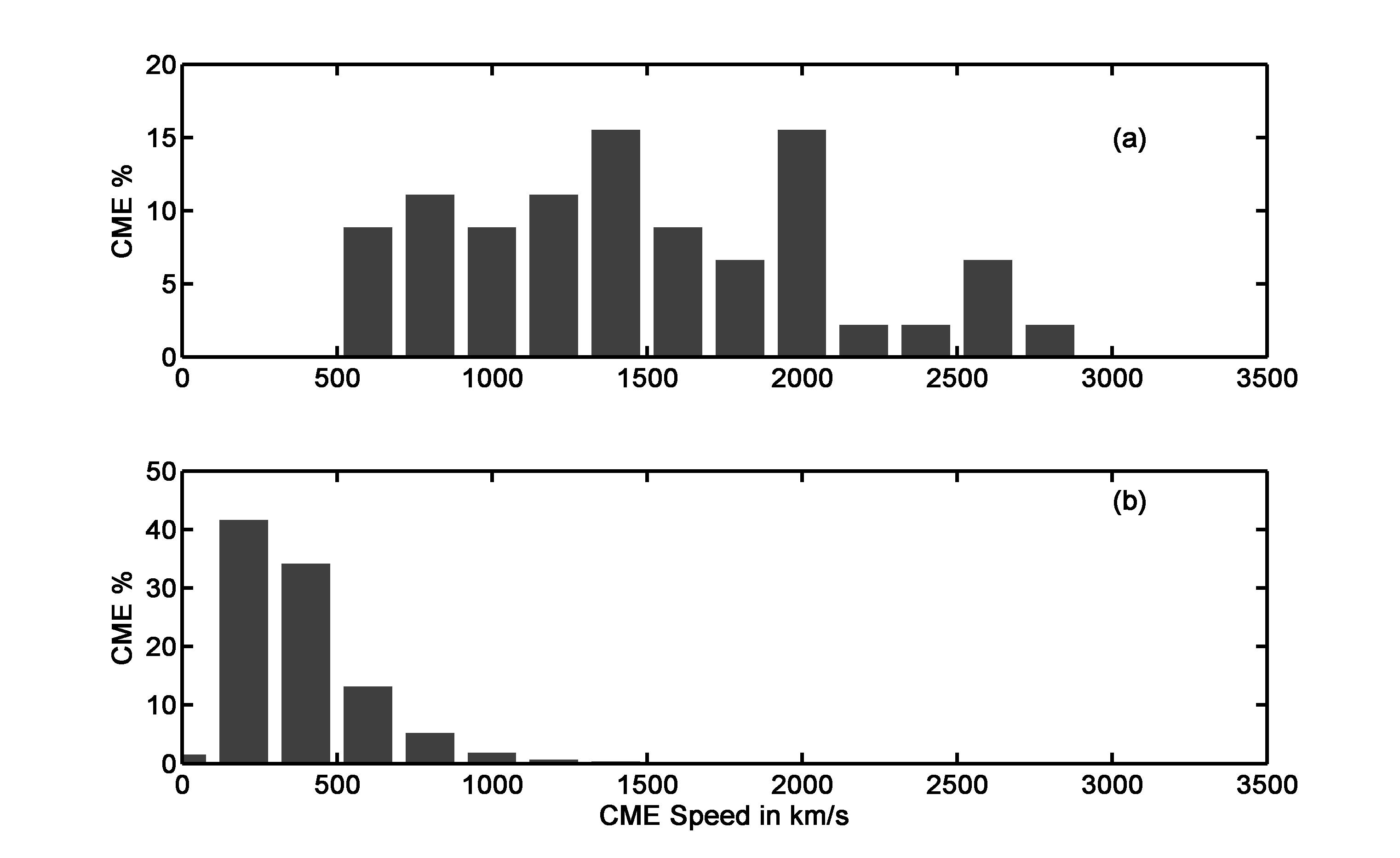} 
  \caption{Distribution (\%) of CME speeds in  km s$^{-1}$. (a) LASCO CMEs associated with compact events. (b)  All LASCO CMEs in the 1998--2012 period for comparison.}
\label{CMEspeed}
 \end{figure}
%-------------------------------------------------
%-------------------------------------------------
 \begin{figure}
\centering
\includegraphics[trim=8.0cm 2.0cm  6.0cm 4.0cm,clip=true,width=\textwidth]{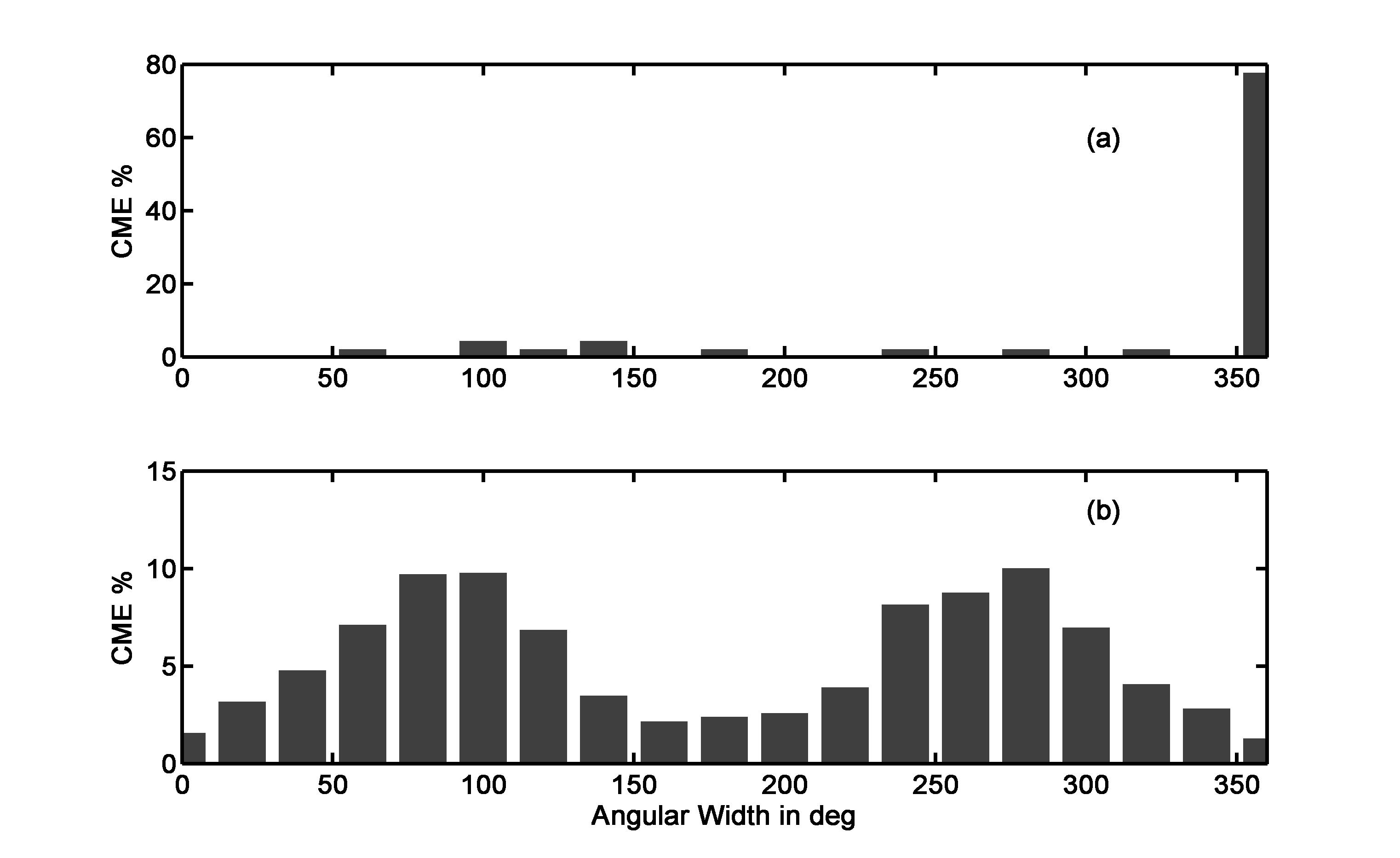} 
  \caption{Distribution (\%) of CME widths. (a)  LASCO CMEs associated with compact events. (b) All LASCO CMEs in the 1998--2012 period for comparison.}
\label{CMEwidth}
 \end{figure}
%-------------------------------------------------

As regards {\bl the CME compact-event} association, \CompactFast ~ of these events were trailing CMEs with speeds $\sim$1400 {\bl km s$^{-1}$} on average with only \CompactSlow ~events having CMEs slower than \mbox{1000 {\bl km s$^{-1}$} } (600--900 {\bl km s$^{-1}$}  range). This  deviates significantly from the LASCO CME distribution {\bl from 1998 to 2011. The} comparison of the distributions is presented {\bl in Figure} \ref{CMEspeed}.  The same CMEs were found, systematically, in the $\sim$360\DG~tail of the width distribution (see Figure \ref{CMEwidth}). These fast and wide CMEs were expected to {\bl transport} the type IV emitting energetic electrons confined within their cavities. \citet{Bain2014} calculated  that the electrons accelerated during the CME initiation or early propagation phase, trapped in the magnetic structure of the CME, do not need to be replenished for times of the order of {\bl four} hours. {\bl This} is consistent with the duration of the {\bl compact} type IV bursts which is, on average, about 106 minutes. Furthermore, \PreCondition ~{\bl compact} events were characterized by a CME preceding by some hours  the associated fast CME along the same path {\bl (similar measured position angle in the LASCO CME catalog)} could  have reduced the propagation drag of the trailing CME. 

Contrary to the discussion in the previous paragraph, {\bl during the interval from 28 October 2003 at 11:30 UT to 29 October at 10:17 UT,} no CME was reported in the SOHO/LASCO catalog, yet two type \IVI~bursts, in close succession, were recorded by the \WIND. These  correspond to entries  28  and 29  in {\bl the attached online table} and represent two {\bl compact} events associated with {\bl M-class} flares. In the {\bl column with remarks} of the CME catalog, at the line corresponding to the 28 October 2003 11:30 halo CME and the associated X17.2 flare, it is noted that "\emph{all images after 13:00 UT, particularly C3, are severely degraded due to the ongoing proton storm}".  The non-detection of CMEs associated with {\bl the events on 28 and 29 May can, therefore,  be due to this fact.}

\subsection{\bl Characteristics  of the \Extended~Extended  Events}\label{CharExtended}

As regards the \Extended  {\bl extended} or {\bl long duration} type \IVI~events, they  seem to need a resupply of the continuum as {\bl they last} from  960 min to 115 hours (examples in {\bl Figures} \ref{DynSpec19990527} and  \ref{DynSpec20020518}); this requirement holds regardless of the intensity and speed of the first associated flare/CME.  Now the  energetic electron sources in the corona manifest themselves as metric--decametric type III and type IV radio bursts. These electron sources need be associated with the lift off and propagation of CMEs as they deform the solar magnetic field providing a propagation path for the energetic electrons and, at the same time, a moving magnetic trap.

In  the examples of {\bl extended}  interplanetary type IV bursts {\bl discussed in Sections} \ref{Over01}, \ref{Over01B}  and \ref{Over01C}  we see the {\bl replenishment} process, mentioned above, {\bl  at work. In all cases}  we have, in addition to the dynamic spectra, a partial coverage with NRH images. The energetic electron sources in the corona manifesting themselves as metric and decametric  type IV radio bursts persist, in the same position, for the duration of each of the interplanetary type IV bursts.  

There are, however,  other possible sources of energetic electrons. Firstly, the  type III bursts; these appear {\bl to extend in the dynamic spectra far} beyond the low frequency limits of the type \IVI~{\bl bursts; therefore,} a mechanism of electron deposition into the type \IVI~is not easily envisaged. The type III--like activity, however, embedded within the type IV continua as part of the type IV fine structure is linked to the type IV energetic population and the corresponding acceleration process. {\bl Another kind of type} III--like activity, are the {\bl micro-type III} bursts which are parts of the {\bl IP storms}. As they are significantly weaker than the standard type III \citep[\bl six orders of magnitude, see ][]{2007_Morioka} they are difficult to detect, especially in the presence of type IIIs but they cannot be ruled out.  

Secondly, the type \IVI~replenishment from the shock accelerated electrons is considered. The type II bursts appear, mostly, piston driven by CMEs and preceding the type IV continuum, interplanetary and coronal, which evolves in their wake, possibly within the CME core. This implies a sort of magnetic isolation from energetic populations {\bl in the vicinity of} the CME bow shock.  This been said, we note that the possibility of acceleration in the low corona by shocks distinct from those preceding the type IV cannot be precluded; the observational confirmation is difficult however as these are often burried in other types of radio activity.

\section{Discussion and Conclusions} \label{discussion}
The present study {\bl is} based on a multi--frequency and multi--instrument study of a sample of  \TotalEvents ~interplanetary type IV (type \IVI) bursts identified from the \WIND~{\bl on line catalog.} The dynamic spectra obtained from the \WIND ~R1 and R2 receivers, in the hectometric frequency range, were combined with metric and decametric dynamic spectra and supplemented with GOES SXR light curves and LASCO CME data. 

In most cases, \Compact~out of \TotalEvents, the extension of a metric--decametric moving  type IV bursts in the hectometric frequency range was found to be associated to a fast  and wide CME (see Section \ref{CharCompact}) capable of driving the embedded  type IV source into the high corona. This type of bursts have duration of about 106 minute, on average; they were dubbed  {\bl compact} type \IVI~bursts. The reduced  {\bl aerodynamic drag} in the wake of a previous CME, along the same propagation path,  appears to increase the probability of the appearance of a type \IVI~burst. This preconditioning of interplanetary space by a previous CME was first proposed, before the discovery of CMEs, by \citet{Caroubalos64} who stated that a disturbance following a preceding  disturbance encounters much more regular conditions than the first. This result corresponds to the effect of CMEs on the structure of the ambient magnetic field and solar wind flow which in turn controls the propagation behavior of trailing CMEs as discussed in a number of publications \citep{2007_Vrsnak&Zic, 2008_Gopalswamy, 2013_Baker&al, 2014_Vrsnak&alShort, 2013_Liu&alshort, 2015_Temmer&Nitta}. The basic argument, in all  cases, is  that a CME may be subject to a {\bl minimal slow down} in the wake of a preceding CME, as it  encounters a {\bl preconditioned} region of depleted ambient plasma density and almost radial magnetic field lines; within this region a reduced {\bl aerodynamic drag}  is expected. The efficiency of this effect increases, possibly,  if the {\bl main} CME  is quite dense \citep[as discussed by~][~based on detailed modelling of a CME propagation]{2015_Temmer&Nitta}. It is also expected that a wide preceding CME {\bl would}  result in a greater drop of the {\bl aerodynamic drag}  compared to a narrow CME along the path of the {\bl main} CME. Further complications may arise due to more than one  CME, preceding the {\bl main} CME, to the projection effects as regards CME paths and speeds and the ambiguities {\bl in CME mass calculations.} Despite these, this report provides qualitative support to the  reduced {\bl aerodynamic drag due to  preconditioning} hypothesis.

{\bl Apart from the preconditioning} of space by a preceding CME, discussed in the previous paragraph, the characteristics of these events were consistent with the {\bl big flare syndrome} since they were mostly associated with medium to large flares and fast CMEs (see Section \ref{CharCompact}).  As regards the small number {\bl (\CompactC~of \Compact) of} events associated with smaller flares we found that either the type IV burst was originating at the solar limb {\bl (numbers 37 and 44 in the attached table) or (numbers 35, 42, and 43)} the origin of the flare was not known; in both cases the flare association was quite uncertain. There were also intense flares within the period of interest, 1998--2012, which did not give type \IVI~bursts. This may be interpreted, at least in part, by the fact that in the \WIND~catalog some are not listed as type IV bursts. On {\bl 20 January 2005}, for example the interplanetary radio signature of the X7.1/2B flare accompanied by a fast ($\approx$900 {\bl km s$^{-1}$}) halo CME was described as {\bl very diffuse}. The same holds for the major solar eruption of 7 March  2012  \citep[X5.4 and X1.3 flares associated with two fast ($>2000$  {\bl km s$^{-1}$}) CMEs, see][]{Patsourakos2013short, Patsourakos2016short}; this is mentioned as {\bl strong intermittent multiple tones.}

{\bl In the \Extended~long duration, or extended}, type \IVI~ bursts, the energetic electron population, which is the type \IVI~source, seems to be replenished from the lower solar corona. This implies the possibility of a connection of the type \IVI~enclosing magnetic structure to low coronal electron accelerators or coronal reservoirs. The NRH images, when available, indicate that these are possibly associated to the high frequency type IV which persists {\bl throughout the duration of the  extended} type \IVI~ burst. 

 A {\bl steady coronal reservoir} of energetic electrons appears to be the metric type IV continuum as most of the type IIIs tend to overshoot the interplanetary type IV. The {\bl micro-type III}  bursts, on the other hand,  may trace the electrons' path from the {\bl coronal reservoir} to the type \IVI. We state, at this point, that the term {\bl coronal reservoir} is used in order to distinguish from the {\bl heliospheric reservoirs} \citep[see~][]{1992_Roelof&al, 2003_Sarris&Malandraki} beyond 1AU. The fact that the {\bl extended} type \IVI~bursts appear to result, cumulatively, from relatively small energetic events, suggests the presence of some type of trapping structure for the exciter energetic electrons. The type \IVI~heliocentric distances, however, are $\sim$ 25--95\RSUN ~which are quite smaller than the 1AU, and beyond, distance of the {\bl heliospheric reservoirs}.

The question of the confinement of the energetic electrons, producing this type of bursts, at heliocentric distances of the order of some tens of \RSUN~ remains open. 

%-------------------------------------------------
\begin{acks}
This research has been partly {\bl cofinanced} by the European Union (European Social Fund ESF) and Greek national funds through the Operational Program {\bl Education and Lifelong Learning} of the National Strategic Reference Framework (NSRF) - Research Funding Program: Thales. Investing in knowledge society through the European Social Fund. The LASCO CME catalog  is generated and maintained at the CDAW Data Centre by NASA and The Catholic University of America in cooperation with the Naval Research Laboratory. SOHO is a project of international cooperation between ESA and NASA. The {\bl \emph{Nan\c cay Radioheliograph} (NRH)} is operated by the Observatoire de Paris and funded by the French research agency CNRS/INSU. {\bl \emph{The Radio Solar Telescope Network}} (RSTN) is a network of solar observatories maintained and operated by the U.S. Air Force Weather Agency. The authors acknowledge the use of the smoothed differentiation filter software by Jianwen Luo. They also thank the anonymous reviewer for valuable comments and useful suggestions.
\end{acks}
\section*{\bl Disclosure of Potential Conflicts of Interest}
{\bl The authors declare that they have no conflicts of interest. }

%-------------------------------------------------

%-------------------------------------------------
\appendix
\section*{\bl Comprehensive Catalog of the Type \IVI~Burst Records}%\label{Catalogue}
{\bl In the table (Table 1), which is attached as online supplementary material,} we provide a summary of the interplanetary type IV bursts recorded by the \WIND~R1 and R2 receivers in the 13.825 MHz--20 {\bl kHz} frequency range, with their associated CMEs and SXR flares in the 1998--2012 period. The coronal extensions of these bursts by the {\bl RSTN, DAM, \emph{ARTEMIS-IV}, CULGOORA, \emph{Hiraiso} and IZMIRAN \emph{Radio Spectrographs}} are included for comparison. 

The headline to each event includes event number, date of observation and characterization of the event as  {\bl compact} or  {\bl extended} following  the classification introduced in Section \ref{Analysis}. Column 1 gives the type of activity, for SXR flares we give the GOES class .{\bl The secondary headline,  CME preceding Main Ejection}, stands for CMEs preceding the {\bl main} CME associated with the event by {\bl approximately} two days along the same path; the path similarity is determined by the comparison of the {\bl position angle (PA)} and width of the preceding CME to the main. In the {\bl Extended Events} the secondary headline {\bl is sometimes absent as} they may originate from a number of energetic events (flares and CMEs) and not from a powerful flare/fast CME with the latter propagating in the wake of preceding ejections (see discussion in {\bl Section} \ref{discussion}).  Columns 2-3 give start, peak, {\bl and end}  of each type of activity in UT and {\bl day month hour minute (\mbox{DD MMM HH:MM})} format  for {\bl start and end. D} indicates that the event extends in time beyond the observation period. The CME start time, {\bl in the second column} is the first C2 appearance, while its  extrapolated lift--off time appears in the  next row as a remark (see below for the description of the remark lines). {\bl In the fourth column,} we give the SXR flare {\bl 1\,--\,8 \AA} ~integrated flux ($\text{F}^{tot}_\text{SXR}$ in {\bl J m$^{-2}$}) and, in the same column, the CME speed ($\text{V}_\text{CME}$) in {\bl km $s^{-1}$}. The location of the flare on the disk  and the CME {\bl measured position angle} (MPA) of the CMEs  with their angular width in parenthesis are given in {\bl in the fifth column. The} SXR flare location is determined from the position of the associated {\bl $\textrm H\alpha$} flare on disk or the {\bl Solar X-ray Imager} of GOES report if available. {\bl In the fifth column,}  we also give the position of the coronal radio bursts when NRH {\bl records} are available. In {\bl In the sixth column, we} give the frequency range of the radio bursts in MHz; the L indicates that the burst extends to lower frequencies, H stands for {\bl high-frequency} extension. 

Comments and remarks, when necessary,  are in separate lines under the description of the activity line.  The comments  include the reporting stations from which the data of each observation was obtained  and  the {\bl classification  of the  \WIND~  and the SOHO/LASCO records and}  the NOAA active region number of the event. For the SXR flares the SXR peak and  the {\bl $\textrm H\alpha$} category when available are reported while for the {\bl CMEs} the extrapolated lift-off time is presented. {\bl Finally, in the comment} lines data gaps, if any, are repoted. 

The reporting observatory or space experiment abbreviations used in {\bl Table 1} are:

\begin{itemize}
\item[ART-4]{{\bl \emph{ARTEMIS-IV}}, Greece}
\item[CUL] {{\bl \emph{Culgoora}}, Australia}
\item[SAG]{{\bl \emph{RSTN: Sagamore Hill}}, Massachusetts, USA}
\item[PAL]{{\bl \emph{RSTN: Palehua}}, Hawaii}
\item[HOL]{{\bl \emph{RSTN:  Holloman}}, New Mexico, USA}
\item[LEA]{{\bl \emph{RSTN:: Learmonth}}, Australia}
\item[SVI]{{\bl \emph{RSTN:: San Vito}}, Italy)}
\item[RAM]{{\bl \emph{Ramey AFB, Puerto Rico}}, USA}
\item[IZM]{{\bl \emph{The radio spectrograph of Izmiran}}}
\item[KANZ]{ {\bl \emph{Kanzelh\"ohe Solar Observatory}}}
\item[MIT]{ {\bl \emph{National Astronomical Observatory of Japan}}, Mitaka}
\item[HiRAS]{{\bl \emph{Hiraiso Radio Spectrograph}}}
\item[DAM]{{\bl \emph{The Nan\c cay Decameter Array}}}
\item[NRH]{{\bl \emph{The Nan\c cay radioheliograph}}}
\item[XFL]{SXR flare from the {\bl \emph{GOES Solar X-ray Imager}} (SXI)}
\item[Gxx]{SXR flare from the {\bl \emph{GOES}}  (for example G08 stands for GOES 08)}
\end{itemize}   

\noindent {All {\bl abbreviations, with the exception of NRH, DAM, and ART-4,} were adopted from the Space Weather Prediction Center\footnote{ftp.swpc.noaa.gov/pub/welcome/stations} {station list}}
     
% %-------------------------------------------------
% %-------------------------------------------------

% Last Modification 8 JUNE 2016
%
\nopagebreak
\begin{longtable}[c]{l ll  lll}

\caption{Type \IVI~Radio Bursts and Associated Activity}
\setlength\tabcolsep{0.01pt} % default value: 5pt
\label{Table02}\\ \nopagebreak
%This is the header for the first page of the table...
\hline
Obs. 				& Start  							&End    			& $\text{F}^{tot}_\text{SXR}$					&Position		& Freq.  \\
						&\multicolumn{2}{l}{Universal Time} 	&	(J m$^{-2}$)											&PA\,-\, Width	& MHz \\
   						&										& 					& $\text{V}_\text{CME}$  							&						& 		 \\
   						&										&						&	(km s$^{-1}$)											&						&				\\
\hline \hline \nopagebreak
\endfirsthead
%This is the header for the remaining page (s) of the table...
\multicolumn{6}{c}{{\tablename} \thetable{}.:Type \IVI ~Radio Bursts and Associated Activity-Continued} \\  
\hline
Obs. 				& Start  							&End    			& $\text{F}^{tot}_\text{SXR}$					&Position		& Freq.  \\
						&\multicolumn{2}{l}{Universal Time} 	&	(J m$^{-2}$)											&PA\,-\, Width	& MHz \\
   						&										& 					& $\text{V}_\text{CME}$  							&						& 		 \\
   						&										&						&	(km s$^{-1}$)											&						&				\\
\hline \hline \nopagebreak
\endhead
%-------------------------------------------------------------------------------------------------------------------------------------------------------------------------------------
   \multicolumn{6}{l}{ {\bl {Compact Event}:  01\,-\,05 March 2012}} \\  \nopagebreak % 01/2012
Type IV			& 05 Mar. 04:15 	& 05 Mar. 07:00 	&    										&  							& 14\,-\,7.0  				\\  \nopagebreak
Type II 			& 05 Mar. 04:00 	& 05 Mar. 12:20 	&    										&  							& 14\,-\,0.4 				\\  \nopagebreak
\multicolumn{6}{c}{ {{Type \IVI  ~and  II \mbox{(F-H)}, by \WIND}}} \\  \nopagebreak
Type IV 			& 05 Mar. 03:33 	& 05 Mar. 08:33 	&    										&  							& 25L\,-\,500H			\\  \nopagebreak
\multicolumn{6}{c}{ {{Coronal  Type IV  by LEA, HiRAS}}} \\  \nopagebreak
Type III 			& 05 Mar. 03:33 	& 05 Mar. 04:48	& 										&  							& 500\,-\,1.0L 				\\  \nopagebreak
\multicolumn{6}{c}{ {{Two Groups (GG) by \WIND,~HiRAS  ~and  LEA}}} \\
\end{longtable}
% \end{landscape}
\end{article}

\begin{thebibliography}{69}
% BibTex style file: spr-mp-sola-cnd.bst (nameyear,cnd), 2011-09-16
\ifx \bisbn   \undefined \def \bisbn  #1{ISBN #1}\fi
\ifx \binits  \undefined \def \binits#1{#1}\fi
\ifx \bauthor  \undefined \def \bauthor#1{#1}\fi
\ifx \batitle  \undefined \def \batitle#1{#1}\fi
\ifx \bjtitle  \undefined \def \bjtitle#1{\textit{#1}}\fi
\ifx \bvolume  \undefined \def \bvolume#1{\textbf{#1}}\fi
\ifx \byear  \undefined \def \byear#1{#1}\fi
\ifx \bissue  \undefined \def \bissue#1{#1}\fi
\ifx \bfpage  \undefined \def \bfpage#1{#1}\fi
\ifx \blpage  \undefined \def \blpage #1{#1}\fi
\ifx \burl  \undefined \def \burl#1{\textsf{#1}}\fi
\ifx \href  \undefined \def \href#1#2{\textsf{#2}}\fi
\ifx \doiurl  \undefined \def
  \doiurl#1{\href{http://dx.doi.org/#1}{\textsf{#1}}}\fi
\ifx \betal  \undefined \def \betal{\textit{et al.}}\fi
\ifx \binstitute  \undefined \def \binstitute#1{#1}\fi
\ifx \bctitle  \undefined \def \bctitle#1{#1}\fi
\ifx \beditor  \undefined \def \beditor#1{#1}\fi
\ifx \bpublisher  \undefined \def \bpublisher#1{#1}\fi
\ifx \bbtitle  \undefined \def \bbtitle#1{\textit{#1}}\fi
\ifx \bedition  \undefined \def \bedition#1{#1}\fi
\ifx \bseriesno  \undefined \def \bseriesno#1{\textbf{#1}}\fi
\ifx \blocation  \undefined \def \blocation#1{#1}\fi
\ifx \bsertitle  \undefined \def \bsertitle#1{\textit{#1}}\fi
\ifx \bsnm \undefined \def \bsnm#1{#1}\fi
\ifx \bsuffix \undefined \def \bsuffix#1{#1}\fi
\ifx \bparticle \undefined \def \bparticle#1{#1}\fi
\ifx \barticle \undefined \def \barticle#1{}\fi
\ifx \botherref \undefined \def \botherref#1{}\fi
\ifx \url \undefined \def \url#1{\textsf{#1}}\fi
\ifx \bchapter \undefined \def \bchapter#1{}\fi
\ifx \bbook \undefined \def \bbook#1{}\fi
\ifx \bcomment \undefined \def \bcomment#1{#1}\fi
\ifx \oauthor \undefined \def \oauthor#1{#1}\fi
\ifx \citeauthoryear \undefined \def \citeauthoryear#1{#1}\fi
\def \endbibitem {}
\ifx \bconflocation  \undefined \def \bconflocation#1{#1} \fi

\bibitem[\protect\citeauthoryear{{Alissandrakis}
  \textit{et~al.}}{2015}]{2015_Alissandrakis}
\begin{barticle}
\bauthor{\bsnm{{Alissandrakis}}, \binits{C.E.}},
\bauthor{\bsnm{{Nindos}}, \binits{A.}},
\bauthor{\bsnm{{Patsourakos}}, \binits{S.}},
\bauthor{\bsnm{{Kontogeorgos}}, \binits{A.}},
\bauthor{\bsnm{{Tsitsipis}}, \binits{P.}}:
\byear{2015},
\bjtitle{\aap}
\bvolume{582},
\bfpage{A52}.
doi:\doiurl{10.1051/0004-6361/201526265}.
\end{barticle}
\endbibitem

\bibitem[\protect\citeauthoryear{{Aschwanden} and
  {Freeland}}{2012}]{2012_Aschwanden&Freeland}
\begin{barticle}
\bauthor{\bsnm{{Aschwanden}}, \binits{M.J.}},
\bauthor{\bsnm{{Freeland}}, \binits{S.L.}}:
\byear{2012},
\bjtitle{\apj}
\bvolume{754},
\bfpage{112}.
doi:\doiurl{10.1088/0004-637X/754/2/112}.
\end{barticle}
\endbibitem

\bibitem[\protect\citeauthoryear{{Aurass}, {Vr{\v s}nak}, and
  {Mann}}{2002}]{Aurass02}
\begin{barticle}
\bauthor{\bsnm{{Aurass}}, \binits{H.}},
\bauthor{\bsnm{{Vr{\v s}nak}}, \binits{B.}},
\bauthor{\bsnm{{Mann}}, \binits{G.}}:
\byear{2002},
\bjtitle{\aap}
\bvolume{384},
\bfpage{273}.
doi:\doiurl{10.1051/0004-6361:20011735}.
\end{barticle}
\endbibitem

\bibitem[\protect\citeauthoryear{{Aurass} \textit{et~al.}}{1999}]{Aurass99}
\begin{barticle}
\bauthor{\bsnm{{Aurass}}, \binits{H.}},
\bauthor{\bsnm{{Vourlidas}}, \binits{A.}},
\bauthor{\bsnm{{Andrews}}, \binits{M.D.}},
\bauthor{\bsnm{{Thompson}}, \binits{B.J.}},
\bauthor{\bsnm{{Howard}}, \binits{R.H.}},
\bauthor{\bsnm{{Mann}}, \binits{G.}}:
\byear{1999},
\bjtitle{\apj}
\bvolume{511},
\bfpage{451}.
doi:\doiurl{10.1086/306653}.
\end{barticle}
\endbibitem

\bibitem[\protect\citeauthoryear{{Bain} \textit{et~al.}}{2014}]{Bain2014}
\begin{barticle}
\bauthor{\bsnm{{Bain}}, \binits{H.M.}},
\bauthor{\bsnm{{Krucker}}, \binits{S.}},
\bauthor{\bsnm{{Saint-Hilaire}}, \binits{P.}},
\bauthor{\bsnm{{Raftery}}, \binits{C.L.}}:
\byear{2014},
\bjtitle{\apj}
\bvolume{782},
\bfpage{43}.
doi:\doiurl{10.1088/0004-637X/782/1/43}.
\end{barticle}
\endbibitem

\bibitem[\protect\citeauthoryear{{Baker} \textit{et~al.}}{2013}]{2013_Baker&al}
\begin{barticle}
\bauthor{\bsnm{{Baker}}, \binits{D.N.}},
\bauthor{\bsnm{{Li}}, \binits{X.}},
\bauthor{\bsnm{{Pulkkinen}}, \binits{A.}},
\bauthor{\bsnm{{Ngwira}}, \binits{C.M.}},
\bauthor{\bsnm{{Mays}}, \binits{M.L.}},
\bauthor{\bsnm{{Galvin}}, \binits{A.B.}},
\bauthor{\bsnm{{Simunac}}, \binits{K.D.C.}}:
\byear{2013},
\bjtitle{Space Weather}
\bvolume{11},
\bfpage{585}.
doi:\doiurl{10.1002/swe.20097}.
\end{barticle}
\endbibitem

\bibitem[\protect\citeauthoryear{{Bastian} \textit{et~al.}}{2001}]{Bastian01}
\begin{barticle}
\bauthor{\bsnm{{Bastian}}, \binits{T.S.}},
\bauthor{\bsnm{{Pick}}, \binits{M.}},
\bauthor{\bsnm{{Kerdraon}}, \binits{A.}},
\bauthor{\bsnm{{Maia}}, \binits{D.}},
\bauthor{\bsnm{{Vourlidas}}, \binits{A.}}:
\byear{2001},
\bjtitle{\apjl}
\bvolume{558},
\bfpage{L65}.
doi:\doiurl{10.1086/323421}.
\end{barticle}
\endbibitem

\bibitem[\protect\citeauthoryear{{Benz}}{1980}]{Benz1980}
\begin{barticle}
\bauthor{\bsnm{{Benz}}, \binits{A.O.}}:
\byear{1980},
\bjtitle{\apj}
\bvolume{240},
\bfpage{892}.
doi:\doiurl{10.1086/158303}.
\end{barticle}
\endbibitem

\bibitem[\protect\citeauthoryear{{Boischot}}{1957}]{Boischot1957}
\begin{barticle}
\bauthor{\bsnm{{Boischot}}, \binits{A.}}:
\byear{1957},
\bjtitle{Acad. Sci. Paris Comptes Rendus}
\bvolume{244},
\bfpage{1326}.
\end{barticle}
\endbibitem

\bibitem[\protect\citeauthoryear{{Boischot}
  \textit{et~al.}}{1980}]{Boischot1980short}
\begin{barticle}
\bauthor{\bsnm{{Boischot}}, \binits{A.}},
\bauthor{\bsnm{{Rosolen}}, \binits{C.}},
\bauthor{\bsnm{{Aubier}}, \binits{M.G.}},
\bauthor{\bsnm{{Daigne}}, \binits{G.}},
\bauthor{\bsnm{{Genova}}, \binits{F.}},
\bauthor{\bsnm{{Leblanc}}, \binits{Y.}},
\bauthor{\bsnm{\etal}}:
\byear{1980},
\bjtitle{\icarus}
\bvolume{43},
\bfpage{399}.
doi:\doiurl{10.1016/0019-1035(80)90185-2}.
\end{barticle}
\endbibitem

\bibitem[\protect\citeauthoryear{{Bougeret}, {Fainberg}, and
  {Stone}}{1983}]{1983_Bougeret}
\begin{barticle}
\bauthor{\bsnm{{Bougeret}}, \binits{J.-L.}},
\bauthor{\bsnm{{Fainberg}}, \binits{J.}},
\bauthor{\bsnm{{Stone}}, \binits{R.G.}}:
\byear{1983},
\bjtitle{Science}
\bvolume{222},
\bfpage{506}.
doi:\doiurl{10.1126/science.222.4623.506}.
\end{barticle}
\endbibitem

\bibitem[\protect\citeauthoryear{{Bougeret}, {Fainberg}, and
  {Stone}}{1984a}]{1984a_Bougeret}
\begin{barticle}
\bauthor{\bsnm{{Bougeret}}, \binits{J.-L.}},
\bauthor{\bsnm{{Fainberg}}, \binits{J.}},
\bauthor{\bsnm{{Stone}}, \binits{R.G.}}:
\byear{1984}a,
\bjtitle{\aap}
\bvolume{136},
\bfpage{255}.
\end{barticle}
\endbibitem

\bibitem[\protect\citeauthoryear{{Bougeret}, {Fainberg}, and
  {Stone}}{1984b}]{1984b_Bougeret}
\begin{barticle}
\bauthor{\bsnm{{Bougeret}}, \binits{J.-L.}},
\bauthor{\bsnm{{Fainberg}}, \binits{J.}},
\bauthor{\bsnm{{Stone}}, \binits{R.G.}}:
\byear{1984}b,
\bjtitle{\aap}
\bvolume{141},
\bfpage{17}.
\end{barticle}
\endbibitem

\bibitem[\protect\citeauthoryear{{Bougeret}
  \textit{et~al.}}{1995}]{Bougeret95short}
\begin{barticle}
\bauthor{\bsnm{{Bougeret}}, \binits{J.-L.}},
\bauthor{\bsnm{{Kaiser}}, \binits{M.L.}},
\bauthor{\bsnm{{Kellogg}}, \binits{P.J.}},
\bauthor{\bsnm{{Manning}}, \binits{R.}},
\bauthor{\bsnm{{Goetz}}, \binits{K.}},
\bauthor{\bsnm{{Monson}}},
\bauthor{\bsnm{\etal}}:
\byear{1995},
\bjtitle{\ssr}
\bvolume{71},
\bfpage{231}.
doi:\doiurl{10.1007/BF00751331}.
\end{barticle}
\endbibitem

\bibitem[\protect\citeauthoryear{{Cane} and
  {Reames}}{1988}]{1988_Cane&Reames_a}
\begin{barticle}
\bauthor{\bsnm{{Cane}}, \binits{H.V.}},
\bauthor{\bsnm{{Reames}}, \binits{D.V.}}:
\byear{1988},
\bjtitle{\apj}
\bvolume{325},
\bfpage{895}.
doi:\doiurl{10.1086/166060}.
\end{barticle}
\endbibitem

\bibitem[\protect\citeauthoryear{{Caroubalos}}{1964}]{Caroubalos64}
\begin{barticle}
\bauthor{\bsnm{{Caroubalos}}, \binits{C.}}:
\byear{1964},
\bjtitle{Ann. Astrophys.}
\bvolume{27},
\bfpage{333}.
\end{barticle}
\endbibitem

\bibitem[\protect\citeauthoryear{{Caroubalos}
  \textit{et~al.}}{2001}]{Caroubalos01short}
\begin{barticle}
\bauthor{\bsnm{{Caroubalos}}, \binits{C.}},
\bauthor{\bsnm{{Maroulis}}, \binits{D.}},
\bauthor{\bsnm{{Patavalis}}, \binits{N.}},
\bauthor{\bsnm{{Bougeret}}, \binits{J.-L.}},
\bauthor{\bsnm{{Dumas}}, \binits{G.}},
\bauthor{\bsnm{{Perche}}, \binits{C.}},
\bauthor{\bsnm{\etal}}:
\byear{2001},
\bjtitle{Exp. Astron.}
\bvolume{11},
\bfpage{23}.
doi:\doiurl{10.1023/A:1011178517069}.
\end{barticle}
\endbibitem

\bibitem[\protect\citeauthoryear{{Delaboudini{\`e}re}
  \textit{et~al.}}{1995}]{Delaboudiniere95short}
\begin{barticle}
\bauthor{\bsnm{{Delaboudini{\`e}re}}, \binits{J.-P.}},
\bauthor{\bsnm{{Artzner}}, \binits{G.E.}},
\bauthor{\bsnm{{Brunaud}}, \binits{J.}},
\bauthor{\bsnm{{Gabriel}}, \binits{A.H.}},
\bauthor{\bsnm{{Hochedez}}, \binits{J.F.}},
\bauthor{\bsnm{{Millier}}, \binits{F.}},
\bauthor{\bsnm{\etal}}:
\byear{1995},
\bjtitle{\solphys}
\bvolume{162},
\bfpage{291}.
doi:\doiurl{10.1007/BF00733432}.
\end{barticle}
\endbibitem

\bibitem[\protect\citeauthoryear{{Fainberg} and {Stone}}{1970a}]{Fainberg70}
\begin{barticle}
\bauthor{\bsnm{{Fainberg}}, \binits{J.}},
\bauthor{\bsnm{{Stone}}, \binits{R.G.}}:
\byear{1970}a,
\bjtitle{\solphys}
\bvolume{15},
\bfpage{222}.
doi:\doiurl{10.1007/BF00149487}.
\end{barticle}
\endbibitem

\bibitem[\protect\citeauthoryear{{Fainberg} and
  {Stone}}{1970b}]{FainbergStone1970SoPh}
\begin{barticle}
\bauthor{\bsnm{{Fainberg}}, \binits{J.}},
\bauthor{\bsnm{{Stone}}, \binits{R.G.}}:
\byear{1970}b,
\bjtitle{\solphys}
\bvolume{15},
\bfpage{433}.
doi:\doiurl{10.1007/BF00151850}.
\end{barticle}
\endbibitem

\bibitem[\protect\citeauthoryear{{Fainberg} and
  {Stone}}{1971}]{FainbergStone1971SoPh}
\begin{barticle}
\bauthor{\bsnm{{Fainberg}}, \binits{J.}},
\bauthor{\bsnm{{Stone}}, \binits{R.G.}}:
\byear{1971},
\bjtitle{\solphys}
\bvolume{17},
\bfpage{392}.
doi:\doiurl{10.1007/BF00150042}.
\end{barticle}
\endbibitem

\bibitem[\protect\citeauthoryear{{Gergely}}{1986}]{Gergely1986}
\begin{barticle}
\bauthor{\bsnm{{Gergely}}, \binits{T.E.}}:
\byear{1986},
\bjtitle{\solphys}
\bvolume{104},
\bfpage{175}.
doi:\doiurl{10.1007/BF00159959}.
\end{barticle}
\endbibitem

\bibitem[\protect\citeauthoryear{{Gopalswamy}}{2004}]{Gopalswamy2005}
\begin{bchapter}
\bauthor{\bsnm{{Gopalswamy}}, \binits{N.}}:
\byear{2004},
In: \beditor{\bsnm{{Gary}}, \binits{D.E.}},
\beditor{\bsnm{{Keller}}, \binits{C.U.}} (eds.)
\bbtitle{Astrophys. Space Sci. Lib.}
\bseriesno{314},
\bfpage{305}.
doi:\doiurl{10.1007/1-4020-2814-8\_15}.
\end{bchapter}
\endbibitem

\bibitem[\protect\citeauthoryear{{Gopalswamy}}{2008}]{2008_Gopalswamy}
\begin{barticle}
\bauthor{\bsnm{{Gopalswamy}}, \binits{N.}}:
\byear{2008},
\bjtitle{J. Atmos. Solar-Terr. Phys.}
\bvolume{70},
\bfpage{2078}.
doi:\doiurl{10.1016/j.jastp.2008.06.010}.
\end{barticle}
\endbibitem

\bibitem[\protect\citeauthoryear{{Gopalswamy}}{2011}]{Gopalswamy2011}
\begin{botherref}
\oauthor{\bsnm{{Gopalswamy}}, \binits{N.}}:
2011,
\textit{Planetary, Solar and Heliospheric Radio Emissions (PRE VII)},
325.
\end{botherref}
\endbibitem

\bibitem[\protect\citeauthoryear{{Gopalswamy}
  \textit{et~al.}}{2009}]{Gopalswamy2009}
\begin{barticle}
\bauthor{\bsnm{{Gopalswamy}}, \binits{N.}},
\bauthor{\bsnm{{Yashiro}}, \binits{S.}},
\bauthor{\bsnm{{Michalek}}, \binits{G.}},
\bauthor{\bsnm{{Stenborg}}, \binits{G.}},
\bauthor{\bsnm{{Vourlidas}}, \binits{A.}},
\bauthor{\bsnm{{Freeland}}, \binits{S.}},
\bauthor{\bsnm{{Howard}}, \binits{R.}}:
\byear{2009},
\bjtitle{Earth Moon Planets}
\bvolume{104},
\bfpage{295}.
doi:\doiurl{10.1007/s11038-008-9282-7}.
\end{barticle}
\endbibitem

\bibitem[\protect\citeauthoryear{{Gorgutsa}
  \textit{et~al.}}{2001}]{Gorgutsa2001}
\begin{barticle}
\bauthor{\bsnm{{Gorgutsa}}, \binits{R.V.}},
\bauthor{\bsnm{{Gnezdilov}}, \binits{A.A.}},
\bauthor{\bsnm{{Markeev}}, \binits{A.K.}},
\bauthor{\bsnm{{Sobolev}}, \binits{D.E.}}:
\byear{2001},
\bjtitle{Astron. Astrophys. Trans.}
\bvolume{20},
\bfpage{547}.
doi:\doiurl{10.1080/10556790108213597}.
\end{barticle}
\endbibitem

\bibitem[\protect\citeauthoryear{{Guidice} \textit{et~al.}}{1981}]{Guidice81}
\begin{bchapter}
\bauthor{\bsnm{{Guidice}}, \binits{D.A.}},
\bauthor{\bsnm{{Cliver}}, \binits{E.W.}},
\bauthor{\bsnm{{Barron}}, \binits{W.R.}},
\bauthor{\bsnm{{Kahler}}, \binits{S.}}:
\byear{1981},
In: \bbtitle{\baas}
\bseriesno{13},
\bfpage{553}.
\end{bchapter}
\endbibitem

\bibitem[\protect\citeauthoryear{{Hillaris}
  \textit{et~al.}}{2011}]{Hillaris2011short}
\begin{barticle}
\bauthor{\bsnm{{Hillaris}}, \binits{A.}},
\bauthor{\bsnm{{Malandraki}}, \binits{O.}},
\bauthor{\bsnm{{Klein}}, \binits{K.-L.}},
\bauthor{\bsnm{{Preka-Papadema}}, \binits{P.}},
\bauthor{\bsnm{{Moussas}}, \binits{X.}},
\bauthor{\bsnm{{Bouratzis}}, \binits{C.}},
\bauthor{\bsnm{\etal}}:
\byear{2011},
\bjtitle{\solphys}
\bvolume{273},
\bfpage{493}.
doi:\doiurl{10.1007/s11207-011-9872-9}.
\end{barticle}
\endbibitem

\bibitem[\protect\citeauthoryear{{Kahler}}{1982}]{1982_Kahler_b}
\begin{barticle}
\bauthor{\bsnm{{Kahler}}, \binits{S.W.}}:
\byear{1982},
\bjtitle{\jgr}
\bvolume{87},
\bfpage{3439}.
doi:\doiurl{10.1029/JA087iA05p03439}.
\end{barticle}
\endbibitem

\bibitem[\protect\citeauthoryear{{Kayser} \textit{et~al.}}{1987}]{1987_Kayser}
\begin{barticle}
\bauthor{\bsnm{{Kayser}}, \binits{S.E.}},
\bauthor{\bsnm{{Bougeret}}, \binits{J.-L.}},
\bauthor{\bsnm{{Fainberg}}, \binits{J.}},
\bauthor{\bsnm{{Stone}}, \binits{R.G.}}:
\byear{1987},
\bjtitle{\solphys}
\bvolume{109},
\bfpage{107}.
doi:\doiurl{10.1007/BF00167402}.
\end{barticle}
\endbibitem

\bibitem[\protect\citeauthoryear{Kerdraon and Delouis}{1997}]{Kerdraon97}
\begin{bchapter}
\bauthor{\bsnm{Kerdraon}, \binits{A.}},
\bauthor{\bsnm{Delouis}, \binits{J.-M.}}:
\byear{1997},
In: \beditor{\bsnm{Trottet}, \binits{G.}} (ed.)
\bbtitle{Coronal Physics from Radio and Space Observations: Proceedings of the
  CESRA Workshop Held in Nouan le Fuzelier, France, 3--7 June 1996},
\bpublisher{Springer},
\blocation{Berlin, Heidelberg},
\bfpage{192}.
\bisbn{978-3-540-68693-4}.
doi:\doiurl{10.1007/BFb0106458}.
\end{bchapter}
\endbibitem

\bibitem[\protect\citeauthoryear{{Klein} and
  {Mouradian}}{2002}]{KleinMouradian02}
\begin{barticle}
\bauthor{\bsnm{{Klein}}, \binits{K.}},
\bauthor{\bsnm{{Mouradian}}, \binits{Z.}}:
\byear{2002},
\bjtitle{\aap}
\bvolume{381},
\bfpage{683}.
doi:\doiurl{10.1051/0004-6361:20011513}.
\end{barticle}
\endbibitem

\bibitem[\protect\citeauthoryear{{Klein} \textit{et~al.}}{2008}]{Klein08}
\begin{barticle}
\bauthor{\bsnm{{Klein}}, \binits{K.}},
\bauthor{\bsnm{{Krucker}}, \binits{S.}},
\bauthor{\bsnm{{Lointier}}, \binits{G.}},
\bauthor{\bsnm{{Kerdraon}}, \binits{A.}}:
\byear{2008},
\bjtitle{\aap}
\bvolume{486},
\bfpage{589}.
doi:\doiurl{10.1051/0004-6361:20079228}.
\end{barticle}
\endbibitem

\bibitem[\protect\citeauthoryear{{Kondo} \textit{et~al.}}{1995}]{Kondo95}
\begin{barticle}
\bauthor{\bsnm{{Kondo}}, \binits{T.}},
\bauthor{\bsnm{{Isobe}}, \binits{T.}},
\bauthor{\bsnm{{Igi}}, \binits{S.}},
\bauthor{\bsnm{{Watari}}, \binits{S.}},
\bauthor{\bsnm{{Tokimura}}, \binits{M.}}:
\byear{1995},
\bjtitle{J.~Commun.~Res.~Lab.}
\bvolume{42},
\bfpage{111}.
\end{barticle}
\endbibitem

\bibitem[\protect\citeauthoryear{{Kontogeorgos}
  \textit{et~al.}}{2006a}]{KontogeorgosShort}
\begin{barticle}
\bauthor{\bsnm{{Kontogeorgos}}, \binits{A.}},
\bauthor{\bsnm{{Tsitsipis}}, \binits{P.}},
\bauthor{\bsnm{{Moussas}}, \binits{X.}},
\bauthor{\bsnm{{Preka-Papadema}}, \binits{G.}},
\bauthor{\bsnm{{Hillaris}}, \binits{A.}},
\bauthor{\bsnm{\etal.}}:
\byear{2006}a,
\bjtitle{\ssr}
\bvolume{122},
\bfpage{169}.
doi:\doiurl{10.1007/s11214-006-7492-8}.
\end{barticle}
\endbibitem

\bibitem[\protect\citeauthoryear{{Kontogeorgos}
  \textit{et~al.}}{2006b}]{Kontogeorgos06short}
\begin{barticle}
\bauthor{\bsnm{{Kontogeorgos}}, \binits{A.}},
\bauthor{\bsnm{{Tsitsipis}}, \binits{P.}},
\bauthor{\bsnm{{Caroubalos}}, \binits{C.}},
\bauthor{\bsnm{{Moussas}}, \binits{X.}},
\bauthor{\bsnm{{Preka-Papadema}}, \binits{P.}},
\bauthor{\bsnm{{Hilaris}}, \binits{A.}},
\bauthor{\bsnm{\etal}}:
\byear{2006}b,
\bjtitle{Exp. Astron.}
\bvolume{21},
\bfpage{41}.
doi:\doiurl{10.1007/s10686-006-9066-x}.
\end{barticle}
\endbibitem

\bibitem[\protect\citeauthoryear{{Kontogeorgos}
  \textit{et~al.}}{2008}]{Kontogeorgos08short}
\begin{barticle}
\bauthor{\bsnm{{Kontogeorgos}}, \binits{A.}},
\bauthor{\bsnm{{Tsitsipis}}, \binits{P.}},
\bauthor{\bsnm{{Caroubalos}}, \binits{C.}},
\bauthor{\bsnm{{Moussas}}, \binits{X.}},
\bauthor{\bsnm{{Preka-Papadema}}, \binits{P.}},
\bauthor{\bsnm{{Hilaris}}, \binits{A.}},
\bauthor{\bsnm{\etal}}:
\byear{2008},
\bjtitle{Measurement}
\bvolume{41},
\bfpage{251}.
doi:\doiurl{doi:10.1016/j.measurement.2006.11.010}.
\end{barticle}
\endbibitem

\bibitem[\protect\citeauthoryear{{Lecacheux}}{2000}]{Lecacheux2000}
\begin{barticle}
\bauthor{\bsnm{{Lecacheux}}, \binits{A.}}:
\byear{2000},
\bjtitle{AGU, Washington DC Geophys, Monograph Ser.}
\bvolume{119},
\bfpage{321}.
\end{barticle}
\endbibitem

\bibitem[\protect\citeauthoryear{{Liu}
  \textit{et~al.}}{2014}]{2013_Liu&alshort}
\begin{barticle}
\bauthor{\bsnm{{Liu}}, \binits{Y.D.}},
\bauthor{\bsnm{{Luhmann}}, \binits{J.G.}},
\bauthor{\bsnm{{Kajdi{\v c}}}, \binits{P.}},
\bauthor{\bsnm{{Kilpua}}, \binits{E.K.J.}},
\bauthor{\bsnm{{Lugaz}}, \binits{N.}},
\bauthor{\bsnm{{Nitta}}, \binits{N.V.}},
\bauthor{\bsnm{\etal}}:
\byear{2014},
\bjtitle{Nature Comm.}
\bvolume{5},
\bfpage{3481}.
doi:\doiurl{10.1038/ncomms4481}.
\end{barticle}
\endbibitem

\bibitem[\protect\citeauthoryear{{Magdaleni{\'c}}
  \textit{et~al.}}{2010}]{Magdalenic2010}
\begin{barticle}
\bauthor{\bsnm{{Magdaleni{\'c}}}, \binits{J.}},
\bauthor{\bsnm{{Marqu{\'e}}}, \binits{C.}},
\bauthor{\bsnm{{Zhukov}}, \binits{A.N.}},
\bauthor{\bsnm{{Vr{\v s}nak}}, \binits{B.}},
\bauthor{\bsnm{{{\v Z}ic}}, \binits{T.}}:
\byear{2010},
\bjtitle{\apj}
\bvolume{718},
\bfpage{266}.
doi:\doiurl{10.1088/0004-637X/718/1/266}.
\end{barticle}
\endbibitem

\bibitem[\protect\citeauthoryear{{Magdaleni{\'c}}
  \textit{et~al.}}{2012}]{Magdalenic2012}
\begin{barticle}
\bauthor{\bsnm{{Magdaleni{\'c}}}, \binits{J.}},
\bauthor{\bsnm{{Marqu{\'e}}}, \binits{C.}},
\bauthor{\bsnm{{Zhukov}}, \binits{A.N.}},
\bauthor{\bsnm{{Vr{\v s}nak}}, \binits{B.}},
\bauthor{\bsnm{{Veronig}}, \binits{A.}}:
\byear{2012},
\bjtitle{\apj}
\bvolume{746},
\bfpage{152}.
doi:\doiurl{10.1088/0004-637X/746/2/152}.
\end{barticle}
\endbibitem

\bibitem[\protect\citeauthoryear{{Morioka}
  \textit{et~al.}}{2007}]{2007_Morioka}
\begin{barticle}
\bauthor{\bsnm{{Morioka}}, \binits{A.}},
\bauthor{\bsnm{{Miyoshi}}, \binits{Y.}},
\bauthor{\bsnm{{Masuda}}, \binits{S.}},
\bauthor{\bsnm{{Tsuchiya}}, \binits{F.}},
\bauthor{\bsnm{{Misawa}}, \binits{H.}},
\bauthor{\bsnm{{Matsumoto}}, \binits{H.}},
\bauthor{\bsnm{{Hashimoto}}, \binits{K.}},
\bauthor{\bsnm{{Oya}}, \binits{H.}}:
\byear{2007},
\bjtitle{\apj}
\bvolume{657},
\bfpage{567}.
doi:\doiurl{10.1086/510507}.
\end{barticle}
\endbibitem

\bibitem[\protect\citeauthoryear{{Nindos} \textit{et~al.}}{2008}]{Nindos08}
\begin{barticle}
\bauthor{\bsnm{{Nindos}}, \binits{A.}},
\bauthor{\bsnm{{Aurass}}, \binits{H.}},
\bauthor{\bsnm{{Klein}}, \binits{K.-L.}},
\bauthor{\bsnm{{Trottet}}, \binits{G.}}:
\byear{2008},
\bjtitle{\solphys}
\bvolume{253},
\bfpage{3}.
doi:\doiurl{10.1007/s11207-008-9258-9}.
\end{barticle}
\endbibitem

\bibitem[\protect\citeauthoryear{{Nindos}
  \textit{et~al.}}{2011}]{2011_Nindos&al}
\begin{barticle}
\bauthor{\bsnm{{Nindos}}, \binits{A.}},
\bauthor{\bsnm{{Alissandrakis}}, \binits{C.E.}},
\bauthor{\bsnm{{Hillaris}}, \binits{A.}},
\bauthor{\bsnm{{Preka-Papadema}}, \binits{P.}}:
\byear{2011},
\bjtitle{\aap}
\bvolume{531},
\bfpage{A31}.
doi:\doiurl{10.1051/0004-6361/201116799}.
\end{barticle}
\endbibitem

\bibitem[\protect\citeauthoryear{{Patsourakos}
  \textit{et~al.}}{2013}]{Patsourakos2013short}
\begin{bchapter}
\bauthor{\bsnm{{Patsourakos}}, \binits{S.}},
\bauthor{\bsnm{{Vlahos}}, \binits{L.}},
\bauthor{\bsnm{{Georgoulis}}, \binits{M.}},
\bauthor{\bsnm{{Tziotziou}}, \binits{K.}},
\bauthor{\bsnm{{Nindos}}, \binits{A.}},
\bauthor{\bsnm{{Podladchikova}}, \binits{O.}},
\bauthor{\bsnm{\etal}}:
\byear{2013},
In: \bbtitle{11th Hellenic Astron. Conf.},
\bfpage{10}.
\end{bchapter}
\endbibitem

\bibitem[\protect\citeauthoryear{{Patsourakos}
  \textit{et~al.}}{2016}]{Patsourakos2016short}
\begin{barticle}
\bauthor{\bsnm{{Patsourakos}}, \binits{S.}},
\bauthor{\bsnm{{Georgoulis}}, \binits{M.K.}},
\bauthor{\bsnm{{Vourlidas}}, \binits{A.}},
\bauthor{\bsnm{{Nindos}}, \binits{A.}},
\bauthor{\bsnm{{Sarris}}, \binits{T.}},
\bauthor{\bsnm{{Anagnostopoulos}}, \binits{G.}},
\bauthor{\bsnm{\etal}}:
\byear{2016},
\bjtitle{\apj}
\bvolume{817},
\bfpage{14}.
doi:\doiurl{10.3847/0004-637X/817/1/14}.
\end{barticle}
\endbibitem

\bibitem[\protect\citeauthoryear{{Pick} and {Vilmer}}{2008}]{Pick08}
\begin{barticle}
\bauthor{\bsnm{{Pick}}, \binits{M.}},
\bauthor{\bsnm{{Vilmer}}, \binits{N.}}:
\byear{2008},
\bjtitle{\aapr}
\bvolume{16},
\bfpage{1}.
doi:\doiurl{10.1007/s00159-008-0013-x}.
\end{barticle}
\endbibitem

\bibitem[\protect\citeauthoryear{{Pohjolainen}, {Hori}, and
  {Sakurai}}{2008}]{Pohjolainen08}
\begin{barticle}
\bauthor{\bsnm{{Pohjolainen}}, \binits{S.}},
\bauthor{\bsnm{{Hori}}, \binits{K.}},
\bauthor{\bsnm{{Sakurai}}, \binits{T.}}:
\byear{2008},
\bjtitle{\solphys}
\bvolume{253},
\bfpage{291}.
doi:\doiurl{10.1007/s11207-008-9260-2}.
\end{barticle}
\endbibitem

\bibitem[\protect\citeauthoryear{{Pohjolainen}, {Khan}, and
  {Vilmer}}{1999}]{Pohjolainen1999a}
\begin{bchapter}
\bauthor{\bsnm{{Pohjolainen}}, \binits{S.}},
\bauthor{\bsnm{{Khan}}, \binits{J.I.}},
\bauthor{\bsnm{{Vilmer}}, \binits{N.}}:
\byear{1999},
In: \beditor{\bsnm{{Wilson}}, \binits{A.}}, \betal (eds.)
\bbtitle{Magnetic Fields and Solar Processes},
\bsertitle{ESA SP-}
\bseriesno{448},
\bfpage{991}.
\end{bchapter}
\endbibitem

\bibitem[\protect\citeauthoryear{{Pohjolainen}
  \textit{et~al.}}{2001}]{Pohjolainen2001short}
\begin{barticle}
\bauthor{\bsnm{{Pohjolainen}}, \binits{S.}},
\bauthor{\bsnm{{Maia}}, \binits{D.}},
\bauthor{\bsnm{{Pick}}, \binits{M.}},
\bauthor{\bsnm{{Vilmer}}, \binits{N.}},
\bauthor{\bsnm{{Khan}}, \binits{J.I.}},
\bauthor{\bsnm{{Otruba}}, \binits{W.}},
\bauthor{\bsnm{\etal}}:
\byear{2001},
\bjtitle{\apj}
\bvolume{556},
\bfpage{421}.
doi:\doiurl{10.1086/321577}.
\end{barticle}
\endbibitem

\bibitem[\protect\citeauthoryear{{Pohjolainen}
  \textit{et~al.}}{2007}]{Pohjolainen07}
\begin{barticle}
\bauthor{\bsnm{{Pohjolainen}}, \binits{S.}},
\bauthor{\bsnm{{van Driel-Gesztelyi}}, \binits{L.}},
\bauthor{\bsnm{{Culhane}}, \binits{J.L.}},
\bauthor{\bsnm{{Manoharan}}, \binits{P.K.}},
\bauthor{\bsnm{{Elliott}}, \binits{H.A.}}:
\byear{2007},
\bjtitle{\solphys}
\bvolume{244},
\bfpage{167}.
doi:\doiurl{10.1007/s11207-007-9006-6}.
\end{barticle}
\endbibitem

\bibitem[\protect\citeauthoryear{{Prestage} \textit{et~al.}}{1994}]{Prestage94}
\begin{barticle}
\bauthor{\bsnm{{Prestage}}, \binits{N.P.}},
\bauthor{\bsnm{{Luckhurst}}, \binits{R.G.}},
\bauthor{\bsnm{{Paterson}}, \binits{B.R.}},
\bauthor{\bsnm{{Bevins}}, \binits{C.S.}},
\bauthor{\bsnm{{Yuile}}, \binits{C.G.}}:
\byear{1994},
\bjtitle{\solphys}
\bvolume{150},
\bfpage{393}.
doi:\doiurl{10.1007/BF00712901}.
\end{barticle}
\endbibitem

\bibitem[\protect\citeauthoryear{{Reiner} \textit{et~al.}}{2006}]{Reiner2006}
\begin{barticle}
\bauthor{\bsnm{{Reiner}}, \binits{M.J.}},
\bauthor{\bsnm{{Kaiser}}, \binits{M.L.}},
\bauthor{\bsnm{{Fainberg}}, \binits{J.}},
\bauthor{\bsnm{{Bougeret}}, \binits{J.-L.}}:
\byear{2006},
\bjtitle{\solphys}
\bvolume{234},
\bfpage{301}.
doi:\doiurl{10.1007/s11207-006-0087-4}.
\end{barticle}
\endbibitem

\bibitem[\protect\citeauthoryear{{Robinson}}{1978}]{Robinson78B}
\begin{barticle}
\bauthor{\bsnm{{Robinson}}, \binits{R.D.}}:
\byear{1978},
\bjtitle{Australian J. Phys.}
\bvolume{31},
\bfpage{533}.
doi:\doiurl{10.1071/PH780533}.
\end{barticle}
\endbibitem

\bibitem[\protect\citeauthoryear{{Robinson}}{1985}]{Robinson85}
\begin{bchapter}
\bauthor{\bsnm{{Robinson}}, \binits{R.D.}}:
\byear{1985},
In: \beditor{\bsnm{{McLean, D.~J.~and~Labrum, N.~R.}}} (ed.)
\bbtitle{Solar Radiophysics: Studies of Emission from the Sun at Metre
  Wavelengths, Cambridge Univ. Press},
\bfpage{385}.
\end{bchapter}
\endbibitem

\bibitem[\protect\citeauthoryear{{Roelof}
  \textit{et~al.}}{1992}]{1992_Roelof&al}
\begin{barticle}
\bauthor{\bsnm{{Roelof}}, \binits{E.C.}},
\bauthor{\bsnm{{Gold}}, \binits{R.E.}},
\bauthor{\bsnm{{Simnett}}, \binits{G.M.}},
\bauthor{\bsnm{{Tappin}}, \binits{S.J.}},
\bauthor{\bsnm{{Armstrong}}, \binits{T.P.}},
\bauthor{\bsnm{{Lanzerotti}}, \binits{L.J.}}:
\byear{1992},
\bjtitle{\grl}
\bvolume{19},
\bfpage{1243}.
doi:\doiurl{10.1029/92GL01312}.
\end{barticle}
\endbibitem

\bibitem[\protect\citeauthoryear{{Sakurai}}{1974}]{Sakurai74}
\begin{barticle}
\bauthor{\bsnm{{Sakurai}}, \binits{K.}}:
\byear{1974},
\bjtitle{Indian J. Radio and Space Phys.}
\bvolume{3},
\bfpage{289}.
\end{barticle}
\endbibitem

\bibitem[\protect\citeauthoryear{{Sarris} and
  {Malandraki}}{2003}]{2003_Sarris&Malandraki}
\begin{barticle}
\bauthor{\bsnm{{Sarris}}, \binits{E.T.}},
\bauthor{\bsnm{{Malandraki}}, \binits{O.E.}}:
\byear{2003},
\bjtitle{\grl}
\bvolume{30},
\bfpage{2079}.
doi:\doiurl{10.1029/2003GL017921}.
\end{barticle}
\endbibitem

\bibitem[\protect\citeauthoryear{{Temmer} and
  {Nitta}}{2015}]{2015_Temmer&Nitta}
\begin{barticle}
\bauthor{\bsnm{{Temmer}}, \binits{M.}},
\bauthor{\bsnm{{Nitta}}, \binits{N.V.}}:
\byear{2015},
\bjtitle{\solphys}
\bvolume{290},
\bfpage{919}.
doi:\doiurl{10.1007/s11207-014-0642-3}.
\end{barticle}
\endbibitem

\bibitem[\protect\citeauthoryear{Usui and Amidror}{1982}]{usui1982}
\begin{botherref}
\oauthor{\bsnm{Usui}, \binits{S.}},
\oauthor{\bsnm{Amidror}, \binits{I.}}:
1982,
\textit{IEEE Trans Biomed Eng.},
686.
doi:\doiurl{10.1109/TBME.1982.324861}.
\end{botherref}
\endbibitem

\bibitem[\protect\citeauthoryear{{Veronig}
  \textit{et~al.}}{2002}]{2002D_Veronig&al}
\begin{barticle}
\bauthor{\bsnm{{Veronig}}, \binits{A.}},
\bauthor{\bsnm{{Temmer}}, \binits{M.}},
\bauthor{\bsnm{{Hanslmeier}}, \binits{A.}},
\bauthor{\bsnm{{Otruba}}, \binits{W.}},
\bauthor{\bsnm{{Messerotti}}, \binits{M.}}:
\byear{2002},
\bjtitle{\aap}
\bvolume{382},
\bfpage{1070}.
doi:\doiurl{10.1051/0004-6361:20011694}.
\end{barticle}
\endbibitem

\bibitem[\protect\citeauthoryear{{Vr{\v s}nak} and {Cliver}}{2008}]{Vrsnak2008}
\begin{barticle}
\bauthor{\bsnm{{Vr{\v s}nak}}, \binits{B.}},
\bauthor{\bsnm{{Cliver}}, \binits{E.W.}}:
\byear{2008},
\bjtitle{\solphys}
\bvolume{253},
\bfpage{215}.
doi:\doiurl{10.1007/s11207-008-9241-5}.
\end{barticle}
\endbibitem

\bibitem[\protect\citeauthoryear{{Vr{\v s}nak} and {{\v
  Z}ic}}{2007}]{2007_Vrsnak&Zic}
\begin{barticle}
\bauthor{\bsnm{{Vr{\v s}nak}}, \binits{B.}},
\bauthor{\bsnm{{{\v Z}ic}}, \binits{T.}}:
\byear{2007},
\bjtitle{\aap}
\bvolume{472},
\bfpage{937}.
doi:\doiurl{10.1051/0004-6361:20077499}.
\end{barticle}
\endbibitem

\bibitem[\protect\citeauthoryear{{Vr{\v s}nak}, {Magdaleni{\'c}}, and
  {Zlobec}}{2004}]{Vrsnak04}
\begin{barticle}
\bauthor{\bsnm{{Vr{\v s}nak}}, \binits{B.}},
\bauthor{\bsnm{{Magdaleni{\'c}}}, \binits{J.}},
\bauthor{\bsnm{{Zlobec}}, \binits{P.}}:
\byear{2004},
\bjtitle{\aap}
\bvolume{413},
\bfpage{753}.
doi:\doiurl{10.1051/0004-6361:20034060}.
\end{barticle}
\endbibitem

\bibitem[\protect\citeauthoryear{{Vr{\v s}nak}
  \textit{et~al.}}{2014}]{2014_Vrsnak&alShort}
\begin{barticle}
\bauthor{\bsnm{{Vr{\v s}nak}}, \binits{B.}},
\bauthor{\bsnm{{Temmer}}, \binits{M.}},
\bauthor{\bsnm{{{\v Z}ic}}, \binits{T.}},
\bauthor{\bsnm{{Taktakishvili}}, \binits{A.}},
\bauthor{\bsnm{{Dumbovi{\'c}}}, \binits{M.}},
\bauthor{\bsnm{{M{\"o}stl}}, \binits{C.}},
\bauthor{\bsnm{\etal}}:
\byear{2014},
\bjtitle{\apjs}
\bvolume{213},
\bfpage{21}.
doi:\doiurl{10.1088/0067-0049/213/2/21}.
\end{barticle}
\endbibitem

\bibitem[\protect\citeauthoryear{White}{2007}]{white2007solar}
\begin{barticle}
\bauthor{\bsnm{White}, \binits{S.M.}}:
\byear{2007},
\bjtitle{Asian J. Phys}
\bvolume{16},
\bfpage{189}.
\end{barticle}
\endbibitem

\bibitem[\protect\citeauthoryear{{Yashiro} \textit{et~al.}}{2004}]{Yashiro04}
\begin{barticle}
\bauthor{\bsnm{{Yashiro}}, \binits{S.}},
\bauthor{\bsnm{{Gopalswamy}}, \binits{N.}},
\bauthor{\bsnm{{Michalek}}, \binits{G.}},
\bauthor{\bsnm{{St.~Cyr}}, \binits{O.C.}},
\bauthor{\bsnm{{Plunkett}}, \binits{S.P.}},
\bauthor{\bsnm{{Rich}}, \binits{N.B.}},
\bauthor{\bsnm{{Howard}}, \binits{R.A.}}:
\byear{2004},
\bjtitle{\jgr}
\bvolume{109}(\bissue{18}),
\bfpage{7105}.
doi:\doiurl{10.1029/2003JA010282}.
\end{barticle}
\endbibitem

\bibitem[\protect\citeauthoryear{{Zharkova} and
  {Kosovichev}}{1999}]{1999_Zharkova&Kosovichev}
\begin{bchapter}
\bauthor{\bsnm{{Zharkova}}, \binits{V.V.}},
\bauthor{\bsnm{{Kosovichev}}, \binits{A.G.}}:
\byear{1999},
In: \beditor{\bsnm{{Vial}}, \binits{J.-C.}},
\beditor{\bsnm{{Kaldeich-Sch{\"u}}}, \binits{B.}} (eds.)
\bbtitle{8th SOHO Workshop: Plasma Dynamics and Diagnostics in the Solar
  Transition Region and Corona},
\bsertitle{ESA SP-}
\bseriesno{446},
\bfpage{755}.
\end{bchapter}
\endbibitem

\end{thebibliography}
\end{document}